\documentclass{aa}
\usepackage{natbib}
\usepackage{amsmath}
\bibpunct{(}{)}{;}{a}{}{,}

\def\init{{0}} 
\def\bx{{\vec{x}}}
\def\bq{{\vec{q}}}

\def\bv{{\vec{v}}}
\def\bu{{\vec{u}}}

\def\bg{{\vec{g}}}

\def\bw{{\vec{w}}}

\def\b0{{\vec{0}}}
\def\bF{{\vec{F}}}
\def\bP{{\vec{P}}}
\def\bX{{\vec{X}}}
\def\bPi{{\vec{\Pi}}}
\def\bT{{\vec{T}}}

\begin{document}

\title{Adhesive gravitational clustering}

\bigskip\medskip

\author{Thomas Buchert\inst{1,2,3} \and Alvaro Dom\'\i nguez\inst{4,5,6}}

\authorrunning{Buchert \& Dom\'\i nguez}

\institute{Arnold Sommerfeld Center for Theoretical Physics, Ludwig--Maximilians--Universit\"at,
Theresienstr. 37, D--80333 M\"unchen, Germany $\;-\;$
\email{buchert@theorie.physik.uni-muenchen.de}
\and
Theory Division, CERN, CH--1211 Gen\`eve 23, Switzerland
\and
Observatoire de la C\^ote d'Azur, Lab. G.D. Cassini
B.P. 4229, F--06304 Nice Cedex 4, France
\and
Max--Planck--Institut f\"ur Metallforschung,
Heisenbergstr.~3, D--70569 Stuttgart, Germany
\and
Institut f\"ur Theor. und Angew. Physik,
Univ.~Stuttgart, Pfaffenwaldring 57, D--70569 Stuttgart, Germany
\and
F\'\i sica Te\'orica, Univ.~Sevilla, Apdo.~1065, E--41080 Sevilla, Spain $\;-\;$
\email{dominguez@us.es}
}

\date{\today}

\abstract{The notion of {\it adhesion} has been advanced for 
the phenomenon of stabilization of large--scale structure 
emerging from gravitational instability of a cold medium. 
Recently, the physical origin of adhesion has been identified:
a systematic derivation of the equations of motion for the
density and the velocity fields leads naturally to the
key equation of the `adhesion approximation' -- however, under 
a set of strongly simplifying assumptions. 
In this work, we provide an evaluation of the current status of
adhesive gravitational clustering and a clear explanation of the
assumptions involved. Furthermore, we propose systematic 
generalizations with the aim to relax some of the simplifying 
assumptions. We start from the general Newtonian evolution 
equations for self--gravitating particles on an expanding 
Friedmann background and recover the 
popular `dust model' (pressureless fluid), which breaks down 
after the formation of density singularities; then we investigate,
in a unified framework,
two other models which, under the restrictions referred to above, 
lead to the `adhesion approximation'. We apply the Eulerian and 
Lagrangian perturbative expansions to these new models and, 
finally, we discuss some non--perturbative results that may serve as 
starting points for workable approximations of non--linear structure 
formation in the multi--stream regime.
In particular, we propose a new approximation that includes, in 
limiting cases, the standard `adhesion model' and the Eulerian as well as
Lagrangian first--order approximations.

}
\maketitle

\section{Gravitational clustering}

The present work aims to push analytical modeling of structure formation
into a regime that may be placed between the formation epoch of 
large--scale structure and the onset of virialization of gravitationally 
bound objects. Phenomenologically,
this regime is characterized by a stabilization of structures that formed
out of gravitational instability of a cold medium. The physical origin of
this {\em adhesive clustering} effect is the balance of gravitational forces and
dynamical stresses in collisionless matter. The latter arise 
from a subsequently establishing multi--stream hierarchy
within collapsing high--density regions. 
Although multi--stream forces tend to 
disperse structures, the resulting effect together with gravity 
tends to stabilize them.
This regime is summarized by the term {\it non--dissipative
gravitational turbulence} advanced by Gurevich \& Zybin (1995). 
The models we investigate are a focus of 
current research, since efforts to simulate Hubble volumes
of the Universe and to understand galaxy halo formation 
are faced within a single approach.
However, we take a more conservative point of view to understand the 
evolution of structure on galaxy cluster scales (for a recent theoretical attempt
to address halo structure within kinetic theory see Ma \& Bertschinger 2004).

The current status of analytical models concerning large--scale structure
{\em formation}  may be centred on Zel'dovich's approximation together with
its foundations (the Lagrangian perturbation theory) and optimizations using
filtering techniques for initial perturbation spectra 
(Zel'dovich 1970, 1973; Shandarin \& Zel'dovich 1989; Buchert 1989, 1992;
Melott 1994; Bouchet et al. 1992, 1995; Sahni \& Coles 1995; 
Buchert 1996; Ehlers \& Buchert 1997). These schemes are capable of modeling 
the evolution of
generic spectra including Cold--Dark--Matter cosmogonies down to 
galaxy cluster scales (Bouchet et al. 1995; Buchert et al. 1994, 
Melott et al. 1995; Wei\ss\ et al. 1997; Hamana 1998). 
Above these scales the optimized Lagrangian schemes roughly reproduce the results
of N--body simulations. Since the exact solution is not known, we seek
agreement between the two modeling techniques used. 
Below these scales, both techniques tend to 
fail as a result of both poorly understood physics and poor resolution power, respectively. 
Agreement between N--body simulations and simple solutions
of Lagrangian perturbation schemes may be considered as
supporting analytical models, but they could as well be considered as a
drawback of N--body simulations given the simplicity of the 
analytical approximations compared with the complexity of non--linear self--gravity.
A deeper understanding of structure formation below galaxy 
cluster scales may require more than improving spatial resolution. 
The use of N--body computing in cosmology is possibly overstated 
as is indicated by the differing results obtained when using different N--body codes
on the scales of interest (Melott et al. 1997; Splinter et al. 1998).
Furthermore, particular features show that caution is 
in order: (i) the validity of the mean field approximation for the
gravitational field strength that is commonly used is
questionable, especially on the scales of galaxy halos;
(ii) the possible emergence of soliton 
states (G\"otz 1988) that arise as a result of the non--linear interaction between
gravity and pressure--like forces; solitons have special
stability properties and  may dominate large--scale
structure at late times; (iii) the behavior of the gravitational field strength
near high--density regions: from a generic integral of the field equations 
(Buchert 1993a, Sect.6.1) one clearly infers proportionality of the field strength and the density,
whereas Lagrangian perturbation schemes (in the regime where they match  
N--body runs) show smooth and moderately increased field strengths 
when crossing high--density regions. In the extreme case of
infinite resolution, Zel'dovich's approximation produces caustics
in the density field, whereas the field strength remains smooth, although
it should blow up at caustics; whether N--body codes treat this 
correctly is questionable in view of the strongly varying fields on the spatial
as well as temporal resolution scales.

In this situation, analytical models that are capable of 
accessing non--linear scales should be further explored. 
This task is not easy,
since --- as explained above --- such models may not be fully guided by comparisons 
with N--body results. 
The pioneering model for adhesive clustering has been built in view of the
shortcomings of Zel'dovich's approximation predicting structure decay
after their formation: a Laplacian forcing has been added {\it ad hoc}
to the evolution equation for the peculiar--velocity. Such a force has the
right property of holding structures together after their formation and,
moreover, the model can be solved exactly in terms of known solutions of
the 3D Burgers' equation (Gurbatov et al. 1989;
Weinberg \& Gunn 1990; Kofman et al. 1990, 1992, 1994; see also: Gurbatov 
et al. 1983, 1985, 1991). 
The phenomenon of non--dissipative gravitational turbulence 
is mimicked by the so--called ``burgulence'' (Frisch \& Bec 2001) and roughhewned into an
``effective viscosity''. However, the performance
of the `adhesion approximation' does not dramatically improve the Lagrangian
perturbation schemes, if the latter are subjected to optimization techniques
as mentioned above: even for a strongly hierarchical power spectrum (power law
index of $n = -1$) the optimized Lagrangian schemes
yield better cross--correlation statistics for the density fields, compared with numerical
simulations (Buchert 1999). However, comparisons have been conducted only on the level 
of large--scale structure (e.g., Sahni et al. 1994 investigate the evolution of voids); 
the role that the `adhesion approximation' 
could play on subcluster scales has not been explored. Other proposals to
analytically model the weakly non--linear regime tried to  
improve the performance for large--scale structure, e.g. 
``freezing'' the initial streamlines of the fluid keeping the velocity potential constant
(so--called `Frozen Flow Approximation', Matarrese et al. 1992), or treating the
gravitational potential as constant (`Frozen Potential Approximation',
Brainerd et al. 1993; Bagla \& Padmanabhan 1994). However, these approximations, 
by their nature, are unable to model a realistic  evolution beyond the epoch 
when the time--dependence of the velocity or the potential
is important, {\it cf.} the systematic comparison of different models carried out 
by Sathyaprakash et al. (1995). 

Lying at the physical core of the problem of covering multi--streaming,
the `adhesion approximation' is a promising model due to a strong support given to it
by Buchert \& Dom\'\i nguez (1998): a 
Laplacian forcing appears naturally in a physical description of adhesive
gravitational clustering and the `adhesion approximation' can be derived
on the basis of a set of --- however, strongly simplifying --- assumptions. 
The Laplacian forcing does not represent a true viscosity, it arises
by combining the gravitational field equations with a dynamical pressure
gradient due to multi--stream stresses. Consequently, momentum is conserved
and energy is not lost: the model is time--reversible.
Multi--streaming implies a reshuffling of energy components: bulk kinetic
energy is transformed into internal kinetic energy, and the gravitational potential 
energy of a high--density region gradually dominates over the tidal interaction 
with its environment, thus effectively isolating the system
and --- in idealized cases --- establishing a
{\em dynamical equation of state} as a relation between gravitational potential
energy and internal kinetic energy.

This reasoning can be made mathematically precise by deriving the
equations of motion for the {\em coarse--grained} (or filtered)
density and velocity fields. Given a smoothing length $L$, the system
is divided into the degrees of freedom of the scales above and below
$L$, respectively. 
The evolution of the smoothed density and velocity fields is in
general dynamically coupled to the degrees of freedom below $L$. 
The standard `dust model' is recovered by
assuming that the dynamical effect of this coupling is negligible.
When the `dust model' develops singularities, this assumption
breaks down, and an improved model has to be used that
accounts for multi--streaming and the coupling to the small--scale
degrees of freedom.  The model introduced by Buchert \& Dom\'\i nguez (1998),
which we call the Euler--Jeans--Newton (`EJN') model, identifies
the main source of the corrections to `dust' as the {\em gravitational
  multi--stream} (GM) effect. This gives rise to a stress tensor of
purely kinetic origin --- as in an ideal gas --- which is added to
the large--scale gravity already considered by the `dust model'. This
stress may be idealized in the `EJN model' phenomenologically as
isotropic, and modelled with a pressure $p(\varrho)$ depending only
on the density $\varrho$, i.e.~{\em a dynamical equation of state}.

Numerical tests (Dom\'\i nguez 2003, Dom\'\i nguez \& Melott
2004) show that a polytropic equation of state, 
$p \propto \varrho^{2-\eta}$ in a certain range of densities is
characteristic for a variety of initial conditions. 
The ``anomaly'' $\eta$
measures the deviations from the naive virial prediction $p \propto \varrho^2$,
depends itself on initial conditions 
($\eta$ ranges from 0 to 1 as the small--scale power is reduced), 
and can be assigned to the fact that
the fluid elements are not isolated and in a stationary state, 
as required when deriving $p \propto \varrho^2$ from the virial theorem. 
An exact integral of the form $p \propto \varrho^{5/3}$ can be 
deduced theoretically under certain strong assumptions
(Buchert \& Dom\'\i nguez 1998).

Recent developments of the `EJN model' concern the 
Lagrangian linear regime (Adler \& Buchert 1999), where solutions can be found by 
extrapolating known results of the Eulerian linear regime; the Lagrangian scheme
has been developed to second order (Morita \& Tatekawa 2001, Tatekawa et al. 2002)
and recently to third order (Tatekawa 2005);
numerical tests employing N--body and hydrodynamical simulations 
have been conducted by Tatekawa (2004a,b).

While the foundations of the `EJN model' can be derived on the basis of 
the velocity moment hierarchy 
of the commonly employed Vlasov--Poisson system,
its practical implementation needs (phenomenological) closure conditions to
truncate the hierarchy.  
An attempt to systematically go beyond phenomenology is the Small--Size Expansion
(`SSE') introduced by Dom\'\i nguez (2000). The mode--mode coupling in the 
equations for the fields smoothed over the scale $L$ is estimated under the 
assumption that the most important contribution to this coupling 
comes from inter--mode coupling at scales $\geq L$.
The systematic nature of this scheme is due to a formal expansion of this 
coupling in powers of $L$. 
The `dust model' is recovered as the lowest order term
(formally setting $L=0$). The corrections also yield a Laplacian forcing
as well as the `adhesion model' under certain simplifying assumptions.
However, the corrections account not only for the GM effect like the
`EJN model', but also for the gravitational influence of the density
inhomogeneities on scales below $L$: the stress tensor has an additional
potential--energy contribution that can be viewed as a correction to 
the mean field approximation for the gravitational field--strength and 
also provides a Laplacian forcing, although
subdominant relative to the one due to the GM effect. This effect may become
dominant if one aims to describe structure evolution on galaxy halo scales.
The stress tensor in the `SSE model' is also a source of 
vorticity by tidal torques and shear stretching. This has been studied recently
with the Eulerian perturbation expansion (Dom\'\i nguez 2002).
 
In the present work the `dust', the `EJN' and the `SSE' models will be
presented in a unified framework, as different closure approximations
to a hierarchy of equations, and
a clearcut description of the weakly non--linear regime of adhesive
gravitational clustering will be offered.  It is expected that the
{\em gravitational multi--stream} process ({\em GM--effect} for short)
as well as stresses arising from corrections to the mean field
approximation may help to establish the stationary configurations that are
usually studied in stellar systems theory, where ``virialized
objects'' are, however, considered as {\em isolated} entities.

\medskip

Sect.~\ref{sec:coarsening} applies the coarse--graining method  
to cosmological structure formation in the phase space of $N$ particles,
resulting in a continuum description which features the Vlasov dynamics
as a subcase. Sect.~\ref{sec:models} describes the 
resulting fully non--linear evolution equations for the density and peculiar--velocity 
fields in Eulerian space and summarizes the assumptions 
that reduce the general problem
to the key equation of the `adhesion approximation'. 
Sect.~\ref{sec:euler} reviews the application of the Eulerian
perturbative expansion and Section~\ref{sec:lagrange} that of the
Lagrangian perturbative expansion.
Sect.~\ref{sec:nonpert} presents new results 
beyond the perturbative regime. Sect.~\ref{sec:summary}
summarizes the results and proposes generalizations of the `adhesion
approximation' as well as future prospects.

\medskip

\underbar{\it Notation}: Eulerian coordinates are
denoted by $\bx$, while Eulerian coordinates that are comoving
with the Hubble flow (i.e. the Lagrangian coordinates of the
background solution) are denoted by $\bq$. In both cases,
Lagrangian coordinates are denoted by $\bX$ (note that Lagrangian
coordinates $\bq$ for the homogeneous solution are regarded as
Eulerian coordinates in an inhomogeneous setting).  
Greek indices refer to particles, while Latin indices refer to  
vector components; a repeated index indicates summation.

\section{Dynamical equations through coarse--graining}
\label{sec:coarsening}

We consider a system of $N$ identical particles evolving under
gravitational forces. In the cosmological context,
these particles are the constituents of the dark matter. The purpose
is to model the dynamical clustering of these particles, which is
supposed to lead to the observed large--scale structure. In the
Newtonian approximation, the equations of motion for the position and velocity 
of the particles in phase space are
\begin{equation}
  \label{newton}
  {\dot\bx}^{(\alpha)} = \bv^{(\alpha)} , \qquad 
  {\dot\bv}^{(\alpha)} = \bg^{(\alpha)} , 
\end{equation}
where the 
gravitational field strength $\bg^{(\alpha)}$ is
determined self--consistently by the Newtonian field 
equations,
\begin{gather} 
  \begin{split}
    \nabla^{(\alpha)} \cdot \bg^{(\alpha)} & = \Lambda - 
    4 \pi G m \sum_{\beta \neq \alpha}^N \delta(\bx^{(\alpha)}-\bx^{(\beta)}) \;, \\
    \nabla^{(\alpha)} \times \bg^{(\alpha)} & = \b0 \;. 
  \end{split}
\label{poisson}
\end{gather}

\noindent
$m$ is the mass of a particle, $G$ denotes Newton's gravitational constant
and $\Lambda$ the cosmological constant.

There is an alternative way of writing these equations. We
define the Klimontovich density in the one--particle phase space as:
\begin{equation}
  \label{f_K}
  f_K (\bx, \bv, t) := \sum_{\alpha=1}^N \delta(\bx-\bx^{(\alpha)}) \;
  \delta(\bv-\bv^{(\alpha)}) \;.
\end{equation}
The equations of motion~(\ref{newton}--\ref{poisson}) lead to a
dynamical equation for this density\footnote{
  We have disregarded the infinities arising from the evaluation of
  the field $\bg (\bx ,t)$ at the particles' positions. These are
  artifacts of using point particles and are treated by smearing
  the particles over an infinitesimal size. For our formal
  manipulations, it suffices with the implicit understanding that the
  infinities can be dropped when they appear.}:
\begin{subequations}
  \label{klim}
  \begin{equation}
    \frac{\partial f_K}{\partial t} + \bv \cdot \frac{\partial f_K}{\partial \bx} + 
    \bg \cdot \frac{\partial f_K}{\partial \bv} = 0 \;,
  \end{equation}
  \begin{equation} 
    \nabla \cdot \bg := \Lambda - 4 \pi G m \int d\bv \; f_K (\bx,\bv, t) \;, 
    \; \nabla \times \bg := \b0 \;. 
  \end{equation}
\end{subequations}
This description of the system is exhaustive and much more detailed
than one is usually interested in. A coarser description may be obtained by
smoothing the Klimontovich density: let ${\cal L}$, ${\cal V}$ denote
respectively the spatial and velocity smoothing scales; then a smooth
phase--space density can be defined as:
\begin{eqnarray}
  \label{f_smooth}
 &f(\bx, \bv, t) :=\nonumber\\
&\int \frac{d\bx'}{{\cal L}^3} \frac{d\bv'}{{\cal V}^3} \; 
  W \left( \frac{\bx-\bx'}{{\cal L}} \right) 
  W \left( \frac{\bv-\bv'}{{\cal V}} \right) 
  f_K (\bx', \bv', t) \;,
\end{eqnarray}
where $W (\cdot)$ is a rotationally symmetric coarsening window
normalized to unity. We can write the evolution equation of this
new density according to Eqs.~(\ref{klim}):
\begin{subequations}
  \label{prevlasov}
  \begin{equation}
    \frac{\partial f}{\partial t} + \bv \cdot \frac{\partial f}{\partial \bx} + 
    \bar\bg \cdot \frac{\partial f}{\partial \bv} = 
    -\frac{\partial}{\partial \bx} \cdot {\bf S}^{(v)}
    -\frac{\partial}{\partial \bv} \cdot {\bf S}^{(g)}\; ,
  \end{equation}
  \begin{equation} 
    \nabla \cdot \bar\bg := \Lambda - 4 \pi G m 
    \int d\bv \; f (\bx,\bv, t) \;,
     \nabla \times \bar\bg := \b0 \; ,
  \end{equation}
with
  \begin{align}
    {\bf S}^{(v)} (\bx, \bv, t) := 
    & \int \frac{d\bx'}{{\cal L}^3} 
    \frac{d\bv'}{{\cal V}^3} \;
    W \left( \frac{\bx-\bx'}{{\cal L}} \right) 
    W \left( \frac{\bv-\bv'}{{\cal V}} \right) \nonumber \\
    & \mbox{} \times (\bv' - \bv) \, f_K (\bx', \bv', t) \;,
  \end{align}
  \begin{align}
    {\bf S}^{(g)} (\bx, \bv, t) := 
    & \int \frac{d\bx'}{{\cal L}^3} 
    \frac{d\bv'}{{\cal V}^3} \;
    W \left( \frac{\bx-\bx'}{{\cal L}} \right) 
    W \left( \frac{\bv-\bv'}{{\cal V}} \right) \; \nonumber \\
    & \mbox{} \times [\bg(\bx', t) - \bar\bg (\bx, t)] \, f_K (\bx', \bv', t).
  \end{align}
\end{subequations}
The field $\bar\bg (\bx,t)$ defined by Eqs.~(\ref{prevlasov}b) is the
gravitational {\em mean field} associated with the phase--space
density $f(\bx,\bv,t)$. The terms ${\bf S}^{(v)}$ and ${\bf S}^{(g)}$
represent the dynamical coupling to the degrees of freedom removed
by the smoothing procedure: ${\bf S}^{(v)}$ accounts for the velocity
dispersion, while ${\bf S}^{(g)}$ represents departures from the mean
field gravity on the smoothing scales. Notice that
Eq.~(\ref{prevlasov}a) is, like the exact Eq.~(\ref{klim}a), a
conservation equation in the one--particle phase space, since the total 
number of particles
$N=\int d\bx \,d\bv f(\bx,\bv,t)= \int d\bx' \,d\bv' f_K (\bx',\bv',t)$. 
 
Eqs.~(\ref{prevlasov}a,b) do not form a closed system of equations
unless the sources ${\bf S}^{(v)}$, ${\bf S}^{(g)}$ can be
represented or approximated as functionals of $f(\bx,\bv,t)$. The form
and success of the approximation depends in general on the initial
conditions and the choice of the coarsening scales ${\cal L}$, ${\cal
V}$. 
A frequently used approximation in many examples of cosmological
and astrophysical interest consists of neglecting small--scale departures 
from the coarse variables, ${\bf S}^{(v)} \approx \b0$, ${\bf
S}^{(g)} \approx \b0$, so that Eqs.~(\ref{prevlasov}a,b) reduce to the
Vlasov--Poisson system of equations for a smooth $f(\bx,\bv,t)$ (e.g.,
Peebles 1980; Binney \& Tremaine 1987). 

However, to describe the formation of cosmological
structures, the continuous phase--space density $f(\bx,\bv,t)$ is still
too detailed, since the observables we are interested in
are the large--scale density and velocity fields. A {\em mass density} and a 
{\em mean fluid velocity} can be defined by the velocity moments of
$f(\bx,\bv,t)$:
\begin{subequations}
  \label{hydrodefs}
  \begin{equation}
    \rho (\bx, t) := m \int d\bv \; f (\bx, \bv, t) =
    \frac{m}{{\cal L}^3} \sum_\alpha W \left(\frac{\bx-\bx^{(\alpha)}}{{\cal L}} \right) \;,
\end{equation}
\begin{equation}
    \rho \bar\bv (\bx, t) := m \int d\bv \; \bv \, f (\bx, \bv, t) = 
    \frac{m}{{\cal L}^3} \sum_\alpha
    W \left(\frac{\bx-\bx^{(\alpha)}}{{\cal L}} \right) \bv^{(\alpha)} ,
  \end{equation}
\end{subequations}
where the last expressions follow from inserting successively the
definitions of $f$, Eq.~(\ref{f_smooth}), and $f_K$, Eq.~(\ref{f_K}).
Eq.~(\ref{hydrodefs}b) defines the Eulerian {\em mean fluid velocity} $\bar\bv$.

The evolution equations for these two fields, expressing mass and
momentum conservation, follow immediately from Eqs.~(\ref{prevlasov})
or from Eqs.~(\ref{newton}--\ref{poisson}):
\begin{subequations}
  \label{hydroeq}
  \begin{equation}
    \frac{\partial \rho}{\partial t} + \nabla \cdot (\rho \bar\bv) = 0 \;,
  \end{equation}
  \begin{equation}
    \frac{\partial \bar\bv}{\partial t} + (\bar\bv \cdot \nabla) \bar\bv = \bar\bg + 
    \frac{1}{\rho} (\vec{\cal F} - \nabla \cdot \vec{\cal P}) \;,
  \end{equation}
  \begin{equation}
    \nabla \cdot \bar\bg := \Lambda - 4 \pi G \rho \;, \qquad
    \nabla \times \bar\bg := \b0 \;. 
  \end{equation}
\end{subequations}
$\bar\bg$ is identical to the gravitational mean field strength defined in
Eq.~(\ref{prevlasov}b). Two new fields have emerged:
\begin{subequations}
  \label{hydrocorrections}
  \begin{eqnarray}
    {\cal F}_i (\bx, t) & := & m \int d\bv \; S_i^{(g)} (\bx, \bv, t) \nonumber \\
    & = & \frac{m}{{\cal L}^3} \sum_\alpha 
    W \left(\frac{\bx-\bx^{(\alpha)}}{{\cal L}} \right)
    [ g_i^{(\alpha)} - {\bar g}_i (\bx, t) ]\;,
  \end{eqnarray}
  \begin{eqnarray*}
    {\cal P}_{ij} (\bx, t) := &
    m \int d\bv & \Bigl\{ [v_i - {\bar v}_i(\bx, t)] 
    [v_j - {\bar v}_j(\bx, t)] \, f (\bx, \bv, t) \\
    & & + [ v_j - {\bar v}_j(\bx, t) ] S_i^{(v)}(\bx, \bv, t) \Bigr\}
  \end{eqnarray*}
  \begin{equation}
     \mbox{} = \frac{m}{{\cal L}^3} \sum_\alpha 
    W \left(\frac{\bx-\bx^{(\alpha)}}{{\cal L}} \right)
    \lbrack v_i^{(\alpha)} v_j^{(\alpha)} - {\bar v}_i {\bar v}_j (\bx, t)\rbrack \;.
  \end{equation}  
\end{subequations}
These new fields, in analogy to ${\bf S}^{(v)}$ and ${\bf S}^{(g)}$,
represent the coupling of the dynamics of the relevant fields $\rho$
and $\bar\bv$ to the configurational and kinetic degrees of freedom
that have been averaged out. 
The symmetric second--rank tensor field $\vec{\cal P}$ accounts for
the velocity dispersion around the mean velocity. As
Eq.~(\ref{hydrocorrections}b) shows, there are two contributions when
working at the level of the smoothed phase--space density
$f(\bx,\bv,t)$, which are subsumed in a single term at the description
level of the point particles.
The vector field $\vec{\cal F}$ contains the deviation from mean field
gravity --- alternatively, $\bar{\bg}$ is the gravitational field
created by the monopole moment of the mass distribution in the
subsystem defined by the window, while $\vec{\cal F}$ is the
contribution from the higher mass--multipoles generated by the small--scale
structure.

The set of hydrodynamic--like Eqs.~(\ref{hydroeq})
can be derived through the intermediate step of a kinetic equation,
Eq.~(\ref{prevlasov}a), or directly from the microscopic particle
dynamics, Eqs.~(\ref{newton}). In the latter case ,${\cal V}$ disappears from the
definitions~(\ref{hydrodefs}, \ref{hydrocorrections}) after
integrating over the velocities, indicating that the precise value of
${\cal V}$ is unimportant for
Eqs.~(\ref{hydroeq}). 

An advantage of the formal derivation through coarse--graining as
presented here is that it makes explicit the smoothing scales ${\cal
  L}$ and ${\cal V}$, which are usually implicit in most applications
and not clearly presented. Although these two scales are arbitrary,
their choice is usually dictated by the physics of the problem at hand.
Typically ${\cal V}=0$, and we know of only one application where
${\cal V} \neq 0$: the numerical codes to solve the Vlasov equation
necessarily have a finite resolution in phase--space and
coarse--graining in velocity is a method to treat the
filamentation problem in velocity space induced by this equation
(Klimas 1987). (In other works (e.g.~Lynden-Bell 1967, Shu 1978)
phase--space coarsening is invoked but the precise value of ${\cal
  V}$, being irrelevant, is not addressed).

The spatial scale ${\cal L}$ is usually chosen in concordance with the
length scale of the phenomena of interest. In the regimes of
validity of the paradigmatic examples of kinetic theory (Boltzmann
equation and Vlasov equation), ${\cal L}$ is much larger than the mean
interparticle distance $\ell$, so that $f(\bx,\bv,t)$ does not resemble the
spiky Klimontovich density $f_K(\bx,\bv,t)$ (Spohn 1991,
and refs.~therein). The Boltzmann equation for dilute gases with
a short range $\sigma$ of interaction 
follows from Eq.~(\ref{prevlasov}a) when the mean field force
$\bar\bg$ is negligible compared to the effect of the term
${\bf S}^{(g)}$, which is dominated by binary ``close encounters'' of
uncorrelated particles.
This holds in the scaling limit $\sigma \rightarrow 0$ with $\ell \sim
\sigma^{2/3}$ (dilute limit) and ${\cal L} \sim \sigma^{0}$ (continuum
limit), and the mean free path, $\lambda \sim
\ell^3/\sigma^2$, remains finite.
The Vlasov equation describes the dynamical evolution of
$f(\bx,\bv,t)$ when the interaction is long--ranged and weak. For the
gravitational interaction, Eqs.~(\ref{poisson}), this holds in the
scaling limit $m \rightarrow 0$ with $\ell \sim m^{1/3}$ (finite mass
density) and ${\cal L} \sim m^{0}$ (continuum limit), so that the
particle distribution is statistically homogeneous on scales below
${\cal L}$ and the mean field $\bar\bg$ dominates over the effect
of ``close encounters'' contained in ${\bf S}^{(g)}$.

Ma \& Bertschinger (2004) study the Klimontovich equation 
for particles moving in a given stochastic large--scale gravitational field, 
intended to be a model of particle evolution
in a galaxy halo environment; after averaging over realizations of the large--scale
gravitational field a new kinetic equation for the ensemble averaged $f$ is obtained
that differs from the Vlasov equation (this ensemble averaging is, in the language of
statistical physics, similar to an average over  `disorder' rather than 
an average over `thermal noise'). 
For comparison, our Eqs.~(\ref{prevlasov}) 
describe the evolution of a single realization of a coarse--grained distribution
(note that the mathematical structure of both averaging approaches are similar with 
reinterpretation of the window function $W (\cdot)$ as a probability density). 
Of course, in addition to coarse--graining one could implement ensemble averaging. 
The addition of a noise term as a model of small--scale degrees of freedom involves 
interesting physics 
(e.g., Berera \& Fang 1994; Barbero et al. 1997; Dom\'\i nguez et al. 1999;
Buchert et al. 1999; Matarrese \& Mohayaee 2002; Antonov 2004).

\section{Dynamical equations for cosmological structure formation}
\label{sec:models}

In this section we apply the method of coarse--graining to the
case of structure formation in cosmology by gravitational instability.
In this case it is convenient to introduce `comoving' Eulerian
coordinates attached to  a homogeneous--isotropic
solution (Friedmann--Lema\^\i tre backgrounds), characterized by the
expansion factor $a(t)$ (and Hubble's function $H=\dot{a} / a$), with
tiny inhomogeneities superimposed as the seeds of structure formation.
In order to subtract the homogeneous--isotropic motion, one defines
comoving positions $\bq^{(\alpha)}$, peculiar--velocities
$\bu^{(\alpha)}$ and gravitational peculiar--accelerations
$\bw^{(\alpha)}$ as follows:
\begin{align}  
\label{comovingtrafo}
\bq^{(\alpha)} & := \frac{1}{a} \bx^{(\alpha)}\;, \nonumber \\
\bu^{(\alpha)} & := \bv^{(\alpha)} - H \bx^{(\alpha)}\;, \\
\bw^{(\alpha)} & := \bg^{(\alpha)} + \frac{4 \pi G \varrho_H - \Lambda}{3} 
\bx^{(\alpha)}\;, \nonumber
\end{align}
where $\varrho_H(t)$ is the homogeneous background density, satisfying
$\dot{\varrho}_H + 3 H \varrho_H =0$. 
Friedmann's equation $3 \frac{\ddot a}{a} =
\Lambda - 4\pi G \varrho_H$ has been used in (\ref{comovingtrafo}). 
In terms of these variables, Eqs.~(\ref{newton}-\ref{poisson})
are: 
\begin{equation}
  \label{comovingeqs}
  {\dot\bq}^{(\alpha)} = \frac{1}{a} \bu^{(\alpha)}\;,\qquad
  {\dot\bu}^{(\alpha)} = \bw^{(\alpha)} - H \bu^{(\alpha)}\;, \nonumber
\end{equation}
\begin{equation}
  \nabla_{\bq}^{(\alpha)} \cdot \bw^{(\alpha)}  =  - 4 \pi G a 
  \left[ \frac{m}{a^3} \sum_{\beta \neq \alpha}^N  
    \delta(\bq^{(\alpha)}-\bq^{(\beta)}) -\varrho_H \right],\;\nonumber
\end{equation}
\begin{equation}
  \nabla_{\bq}^{(\alpha)} \times \bw^{(\alpha)}  = \b0 \;.
\end{equation}

We can now repeat the derivations of \S\ref{sec:coarsening}. We define
a {\em mass density} and a {\em mean peculiar--\-velocity} as follows:
\begin{subequations}
  \label{defs}
  \begin{eqnarray}
    &&\varrho (\bq, t) := \frac{m}{(a L)^3} \sum_\alpha W 
    \left(\frac{\bq-\bq^{(\alpha)}}{L} \right) \;,
  \\
    &&\varrho \bar\bu (\bq, t) := \frac{m}{(a L)^3} \sum_\alpha 
    W \left(\frac{\bq-\bq^{(\alpha)}}{L} \right) \bu^{(\alpha)} \;,
  \end{eqnarray}
\end{subequations}
where $L={\cal L}/a$ is the (comoving) coarsening length. This length is in
principle arbitrary, its use being motivated by the features of
cosmological structure formation one is interested in. 
In general, one should demand that
$L$ will be much larger than the (comoving) mean interparticle
distance, since dark matter discreteness is cosmologically irrelevant.
In the rest of the paper, when we speak about ``small/large scales'', we
use $L$ to set the comparison scale.

The equations obeyed by these fields can be derived as before. Note that
in the following we  drop the overbar used to distinguish
the microscopic velocities and accelerations from the mean velocities and
mean field strengths; we 
always refer to the following equations, so that we no longer need this
distinction (the time--derivative is taken at constant $\bq$ {\em
  and} $L$ in these equations; see Appendix~\ref{app:comoving} for
the relation to Eqs.~(\ref{hydroeq})):
\begin{subequations}
  \label{momenteq}
  \begin{equation}
    \frac{\partial \varrho}{\partial t} + 3 H \varrho + \frac{1}{a} 
    \nabla_\bq \cdot (\varrho \bu) = 0 \;,
  \end{equation}
  \begin{equation}
    \frac{\partial \bu}{\partial t} + \frac{1}{a} (\bu \cdot \nabla_\bq) \bu + H \bu = 
    \bw + \frac{1}{\varrho} \left( \bF - \frac{1}{a} \nabla_\bq \cdot 
    \bPi \right) \;,
  \end{equation}
  \begin{equation}
    \nabla_\bq \cdot \bw := - 4 \pi G a (\varrho - \varrho_H) \;, 
  \end{equation}
  \begin{equation}
    \nabla_\bq \times \bw := \b0 \;. 
  \end{equation}
\end{subequations}
Here, $\bw$ (with overbar omitted) is the mean field strength of the peculiar--gravity, 
and we have
the ``peculiar counterparts'' to the fields~(\ref{hydrocorrections}):
\begin{subequations}
  \label{corrections}
  \begin{equation}
    F_i (\bq, t) := \frac{m}{(a L)^3} \sum_\alpha W 
    \left(\frac{\bq-\bq^{(\alpha)}}{L} \right)
    [ w_i^{(\alpha)} - {w}_i (\bq, t) ]\;,
  \end{equation}
  \begin{equation}
    \Pi_{ij} (\bq, t) := \frac{m}{(a L)^3} \sum_\alpha W 
    \left(\frac{\bq-\bq^{(\alpha)}}{L} \right)
    [ u_i^{(\alpha)} u_j^{(\alpha)} - {u}_i {u}_j (\bq, t) ]\; .
  \end{equation}
\end{subequations}
If one computes dynamical equations for the new fields $\bF$ and
$\bPi$, one obtains further new fields, and so on. In order for
Eqs.~(\ref{momenteq}) to become useful, one has to close this infinite
hierarchy by resorting to physically motivated approximations that yield
the fields $\bF$ and $\bPi$ as functionals of $\varrho$ and $\bu$. That a successful
closure exists is not guaranteed and this usually depends on the choice
of the scale $L$ and the initial and boundary conditions. In the
context of fluids dominated by short--range interactions, there is a
set of closed hydrodynamic equations when the fields $\varrho$ and
$\bu$ vary spatially only over length scales much larger than any
microscopic length (mean free path, correlation length, etc.). In such
cases, one must take $L$ to lie between these microscopic lengths and
the scale of variation of the fields. Closure is achieved with the
assumption of local quasi--equilibrium: the coarsening cells of size
$\sim L$ are idealized as systems in almost thermal equilibrium, the
corrections to equilibrium being proportional to the relatively small
spatial gradients of the fields\footnote{Notice that this closure
  requires one to consider the energy or, equivalently, the
  thermodynamic temperature as another fundamental field on equal
  footing with $\varrho$ and $\bu$.}. In this manner one obtains the
Navier--Stokes and Fourier equations. The existence of a {\em
  hydrodynamic regime} where these equations hold can be supported
theoretically for dilute gases described by the Boltzmann equation by
means of the Chapman--Enskog expansion (Chapman \& Cowling 1991); for
dense fluids, the hydrodynamic regime has the status of an experimental fact.

In the context of cosmological structure formation one cannot follow
this argument, since the notion of thermal equilibrium is not
well--defined. We instead discuss three closures:

\newcounter{model}
\begin{list}{\arabic{model}.}{\usecounter{model} \leftmargin=0pt \itemsep=5mm \itemindent=5mm}
\item {\bf The `dust model'}: one assumes that the large--scale dynamics
  is dominated by the large--scale forces and neglects the coupling to
  the small scales altogether: one sets $F_i \equiv 0$, $\Pi_{ij,j}
  \equiv 0$ in Eqs.~(\ref{momenteq}), and obtains the Euler--Newton system for
  the peculiar--fields:
  \begin{subequations}
    \label{dust}
    \begin{equation}
      \frac{\partial \varrho}{\partial t} + 3 H \varrho + \frac{1}{a} 
      \nabla_\bq \cdot (\varrho \bu) = 0 \;,
    \end{equation}
    \begin{equation}
      \frac{\partial \bu}{\partial t} + \frac{1}{a} (\bu \cdot \nabla_\bq) \bu + H \bu = \bw \;,
    \end{equation}
    \begin{equation}
      \nabla_\bq \cdot \bw = - 4 \pi G a (\varrho - \varrho_H)\;, 
    \end{equation}
    \begin{equation}
      \nabla_\bq \times \bw = \b0 \;.
    \end{equation}
  \end{subequations}
  This system of equations has a long history and has been thoroughly studied 
  in the cosmological context (see, e.g., Peebles 1980; Sahni \& Coles 1995; Ehlers \& Buchert
  1997; Bernardeau et al. 2002). Coming from
  the context of hydrodynamics, the matter model `dust' is also called a
  `pressure--less fluid'; from our reasoning it is clear that this model is characterized 
  by the assumptions that (i) small--scale inhomogeneities are irrelevant, so that mean field 
  gravity is dominant, $F_i \approx 0$, and
  (ii) multi--streaming is absent and small--scale kinetic degrees of freedom 
  are subdominant too, so
  that ``pressure'' drops, $\Pi_{ij} \approx 0$.\footnote{According to Eq.~(\ref{momenteq}b),
  the same model is recovered if $F_i - a^{-1} \Pi_{ij,j} \approx 0$, i.e., 
  when the corrections to the mean field are counterbalanced by velocity
  dispersion --- we have not explored the physical meaning of such a
  balance condition.}
  
  The `dust model' was originally applied to the `top--down scenario' of structure formation,
  e.g., `Hot Dark Matter cosmogony',
  in which large--scale density inhomogeneities grow while the small scales
  remain homogeneous. The model, however, turned out to be
  a relatively good description also for the `bottom--up scenario', e.g., `Cold Dark Matter
  cosmogony' (Pauls \& Melott 1995), in which the small scales are always
  strongly inhomogeneous. 
  Strictly speaking, the `dust model' is not properly defined if there
  is too much initial small--scale power, because of the emergence of
  singularities at arbitrarily short times, so that a small--scale
  cutoff is required (this becomes manifest in formal perturbative
  expansions too (Valageas 2002; Bernardeau et al. 2002)).
  Our derivation of the 
  `dust model' clearly demonstrates that the smoothing length $L$ is indeed a defining
  ingredient of the model. However, this fact has not been properly
  emphasized in the literature --- the smoothing length is likely
  irrelevant in the `top--down scenario', where the initial conditions
  have a built--in length of smoothness (free--flight scale of the dark matter
  particles), but this is not so in a `bottom--up scenario': in this
  latter case, it was found that a much better
  agreement with N--body simulations is achieved if the initial
  conditions are first smoothed, e.g., the ``Truncated Zel'dovich
  Approximation'' (Coles et al. 1993).
    
  The `dust model' correctly describes many features of the formation of
  structure by gravitational instability. It has also an important
  shortcoming, which is the focus of the present work, namely that it
  generates caustics, i.e., density singularities, as well as {\em
  multi--streaming} regions ({\em shell--crossing}), where the velocity
  field is multi--valued.  This points to a breakdown of the
  approximation, so that the term $F_i-a^{-1} \Pi_{ij,j}$ in
  Eq.~(\ref{momenteq}b) is no longer negligible.  
  In the present work we address
  two approximations beyond the `dust model' in which this term is
  modelled as a function of $\varrho$ and $\bu$.  

\item {\bf The Euler--Jeans--Newton (`EJN') model}: motivated by
  hydrodynamics and the theory of stellar systems, we have the
  following phenomenological closure: the corrections to mean field gravity are
  neglected, and the velocity dispersion
  is approximated by an isotropic tensor field which, in addition, is commonly
  modelled as a function of the local density: 
  \begin{equation}
    \label{EJNcorrections}
    F_i = 0\;, \qquad \Pi_{ij} = p(\varrho) \delta_{ij}\; ,
  \end{equation}
  with a given function $p(\varrho) > 0$, a kinetic pressure.
  (Appendix~\ref{app:eqstate}
  discusses some examples of the functional dependence $p(\varrho)$.)
  This velocity dispersion can be a consequence of the
  multi--streaming or it can be already present in the initial
  conditions.
  This model is applicable to situations in which 
  deviations from the mean field can be ignored. For cosmological structure formation
  in a `top--down evolution', it would be safe to ignore deviations from mean field, since 
  then the scales below the collapsing ones are relatively homogeneous 
  (e.g., due to the free--flight of the dark matter particles as in a 
  `Hot--Dark--Matter cosmogony'), and the main 
  correction to `dust' is due to velocity dispersion.
  On the other hand, if a scenario that entails
  `bottom--up' structure formation is investigated realistically, especially
  down to galaxy halo scales, then the scales below the
  collapsing ones are very inhomogeneous due to the previously formed
  clusters, and both velocity dispersion and stresses due to deviations 
  from mean field gravity can be expected to be relevant.

  Given the phenomenological character of the
  approximation~(\ref{EJNcorrections}), one can think of the relationship 
  $p(\varrho)$ as ``templates'' modeling the overall effect of both velocity dispersion
  and departure from mean field gravity as long as the mathematical
  analysis of Eqs.~(\ref{EJN}) is involved. The only exact constraint
  required by this interpretation is that $\bF$ can be written as the
  divergence of a stress tensor. This is the case  in the other
  closure ansatz to be discussed (see Eq.~(\ref{Fstress})).

\vspace{6pt}

  Inserting the ansatz for $F_i$ and $\Pi_{ij}$ into
  Eq.~(\ref{momenteq}b), one gets (${\Delta}_\bq$ is the
  Laplacian operator in the variable $\bq$):
  \begin{subequations}
    \label{EJN}
    \begin{equation}
      \frac{\partial \varrho}{\partial t} + 3 H \varrho + \frac{1}{a} \nabla_\bq 
      \cdot (\varrho \bu) = 0 \;,
    \end{equation}
    \begin{eqnarray}
      \frac{\partial \bu}{\partial t} + \frac{1}{a} (\bu \cdot \nabla_\bq) \bu + H \bu & = &
      \bw  - \frac{p'(\varrho)}{a \varrho} \nabla_\bq \varrho \nonumber \\ 
      & = & \bw + \frac{\lambda^2 (\varrho)}{a^2} {\Delta}_\bq \bw \;,
    \end{eqnarray}
    \begin{equation}
      \nabla_\bq \cdot \bw = - 4 \pi G a (\varrho - \varrho_H) , \qquad
    \end{equation}
    \begin{equation}
      \nabla_\bq \times \bw = \b0 \;,
    \end{equation}
  \end{subequations}
  where we have defined a density--dependent length (see
  Eq.~(\ref{deltalinear}) for the relation to Jeans' length):
  \begin{equation}
    \label{lambda}
    \lambda (\varrho) := \left[ \frac{p'(\varrho)}{4 \pi G \varrho} 
    \right]^{1/2} \;.
  \end{equation}
  
  Eqs.~(\ref{EJN}a,b) are Euler's equations for a
  compressible fluid, Eq.~(\ref{EJN}b) is known as Jeans' equation in
  stellar systems theory. This model was proposed by Buchert \&
  Dom\'\i nguez (1998) starting from Vlasov's equation for the
  one--particle distribution function, so that corrections to mean
  field gravity were neglected from the outset and only velocity
  dispersion remained. In that work it was also suggested that the
  phenomenological ansatz Eq.~(\ref{EJNcorrections}) could be replaced
  by a different closure condition leading to $p \propto
  \varrho^{5/3}$ (see Appendix~\ref{app:eqstate}).

  The case of isotropic stresses formally covers the hydrodynamics of
  a perfect fluid or, as common in studies of stellar systems,
  polytropic models. However, there is no compelling reason for
  isotropic stresses in collisionless systems before the onset of
  virialization.  Even ``virialized'' systems will in general maintain
  an anisotropic component. At the moment, we consider the isotropy
  assumption as a good working hypothesis to understand the role of the
  interaction between multi--stream forces and self--gravity: 
  the velocity dispersion ellipsoid is approximated by a sphere.
  (Interesting considerations of the influence of the anisotropic part
  have been reported by Maartens et al. 1999; see also Barrow \&
  Maartens 1999.)

\item {\bf The Small--Size Expansion (`SSE')}: this is a method proposed
  by Dom\'\i nguez (2000, 2002) to formalize the notion that
  the coupling to the small scales is in some sense weak.
  One argues that the corrections $F_i$ and $\Pi_{ij,j}$ are determined
  mainly by the largest scales contributing to them, i.e., those close
  to $L$. For instance, in the bottom--up scenario,
  the large--scale dynamics is sensitive mainly to the
  motion of the most recently formed clusters as a whole, and not so much to
  their internal structural details due to the trapped particles, so that one takes $L \sim $
  typical size of these clusters (which play the role of effective
  particles of size $\sim L$). 
  The mathematical implementation of this idea leads to a formal
  expansion of $F_i$ and $\Pi_{ij}$ in powers of the smoothing
  length\footnote{The denomination ``large--scale expansion''
  used in Dom\'\i nguez (2000) may lead to confusion.} $L$ (see
  Appendix~\ref{app:sse} for an outline of the derivation):
  \begin{subequations}
    \label{SSEcorrections}
    \begin{align}
      F_i & \rightarrow B L^2 (\partial_k \varrho) (\partial_k w_i) + 
      {\cal O}(L^4) \;, \\
      \Pi_{ij} & \rightarrow B L^2 \varrho (\partial_k u_i) (\partial_k u_j) 
      + {\cal O}(L^4) \;,
    \end{align}
  \end{subequations}
  where $B>0$ is a dimensionless constant of order unity depending on
  the shape of the smoothing window $W(\cdot)$, Eq.~(\ref{eq:B}).  The
  correction to mean field, being proportional to $\partial_k w_i$, is
  approximated by the tidal forces induced by the mean field gravity.
  With Eqs.~(\ref{momenteq}c,d), it can be written also as the
  divergence of a second--rank tensor:
  \begin{equation}
    \label{Fstress}
    F_i \rightarrow B L^2 \partial_j (\varrho \, \partial_j w_i + 2 \pi G a \varrho^2 \delta_{ij}) 
    + {\cal O}(L^4) \;.
  \end{equation}
  The velocity dispersion tensor is 
  represented through 
  the kinematical parts of the peculiar--velocity gradient: expansion
  rate $\theta:=(1/a) \nabla_\bq \cdot \bu$, vorticity $\boldsymbol{\omega}:=(1/a) 
 \nabla_\bq \times \bu$, and shear $\sigma_{ij}$:
  \begin{equation}
    \label{nablau}
    \frac{1}{a} \partial_i u_j = \frac{1}{3} \theta \delta_{ij} + \sigma_{ij} + 
    \frac{1}{2} \varepsilon_{ijk} \omega_{k} \; ,
  \end{equation}
  \begin{align}
    \Pi_{ij} \rightarrow a^2 B L^2 \varrho \, \biggl[ & \frac{1}{9} \theta^2 \delta_{ij} + 
      \frac{2}{3} \theta \sigma_{ij} + \sigma_{ik} \sigma_{kj} - \mbox{} \nonumber \\
    & \frac{1}{2}(\varepsilon_{ikn} \sigma_{kj} + \varepsilon_{jkn} \sigma_{ki})\omega_n 
      + \mbox{} \nonumber \\
      & \frac{1}{4} (\omega^2 \delta_{ij} - \omega_i \omega_j) \biggr] + {\cal O}(L^4) .
  \end{align}
  In the context of the `SSE model', 
  we recover the `dust model' to order ${\cal O}(L^0)$. To order ${\cal O}(L^2)$, we obtain the
  following equations: 
  \begin{subequations}
    \label{sse}
    \begin{equation}
      \frac{\partial \varrho}{\partial t} + 3 H \varrho + \frac{1}{a} 
      \nabla_\bq \cdot (\varrho \bu) = 0 \;,
    \end{equation}
    \begin{align}
      \frac{\partial \bu}{\partial t} & + \frac{1}{a} (\bu \cdot \nabla_\bq) \bu + H \bu = \bw + 
      \mbox{} \\
      & \frac{B L^2}{\varrho} \left\{ (\nabla_\bq \varrho \cdot \nabla_\bq) \bw
        - \frac{1}{a} \nabla_\bq \cdot [\varrho (\partial_k \bu) (\partial_k \bu)] \right\}\;, \nonumber
    \end{align}
    \begin{equation}
      \nabla_\bq \cdot \bw = - 4 \pi G a (\varrho - \varrho_H) \;, \qquad
    \end{equation}
    \begin{equation}
      \nabla_\bq \times \bw = \b0 \;.
    \end{equation}
  \end{subequations}
  
  The expansions~(\ref{SSEcorrections}) are formally exact. The
  assumption underlying the `SSE' is that the first few terms provide
  an accurate description of the dynamical effect of the mode--mode
  coupling for the problem at hand. For comparison, this is not true in the 
  hydrodynamic regime of fluids
  dominated by short--ranged interactions, where the high--order terms
  in the expansions have to add up to yield the usual $L$--independent
  hydrostatic pressure and viscous force.
  
  The closure assumption~(\ref{SSEcorrections}b) for the velocity
  dispersion has been also used to model the influence of
  subresolution degrees of freedom of the velocity field in
  ``Large--Eddy Simulations'' of turbulent flow (Pope 2000). It has
  been termed the `gradient model' (see, e.g., Vreman et al.~1997) and it
  was introduced as part of the `Clark model' (Clark et al.~1979) (a
  model which includes an extra additive term in the expression
  for $\Pi_{ij}$).
\end{list}

Both the `EJN' and the `SSE' models provide an extension of the `dust model'
with the potential to improve on it by preventing the formation of
singularities. The `EJN model' relies on a phenomenological assumption
concerning the kinetic stress $\Pi_{ij}$, but has the advantage that
the resulting equations are well established. In favor of the `SSE'
model is the fact that the corrections to `dust' can be obtained
systematically beyond phenomenology, but the mathematical and physical
status of the resulting equations is little explored yet.

\medskip

\subsection{Where do these equations hide the `adhesion approximation'?}
\label{sec:adhesion}

In this subsection we get a first picture of the effect of the
corrections to the `dust model'.
In the `adhesion model' (Gurbatov et al.  1989), the evolution
of `dust' is modelled with Zel'dovich's approximation (Zel'dovich
1970, 1973), a powerful and mathematically manageable description of
the exact evolution (first--order Lagrangian perturbation solution,
see \S\ref{sec:dustLang}). Accordingly, one assumes in
Eq.~(\ref{dust}b) the proportionality of peculiar--velocity and
\mbox{--acceleration} fields ({\it cf.} Peebles 1980; Buchert 1989, 1992;
Bildhauer \& Buchert 1991; Kofman 1991; Buchert 1993b; Vergassola et al. 1994):
\begin{equation}
\bw = F(t) \bu , \qquad F(t) = 4\pi G\varrho_H \frac{b(t)}{{\dot b}(t)} \;\;\;,
\label{parallelism}
\end{equation}
where $b(t)$ is identical to the growing--mode solution of the Eulerian
linear theory of gravitational instability for `dust' (i.e., it
solves the equation $\ddot{b} + 2 H \dot{b} - 4 \pi G \varrho_{H} b =
0$). Notice that this ``slaving'' assumption implies an irrotational flow: $\nabla_{\bq}
\times \bu =\b0$. Changing the temporal variable from $t$ to $b$ and
defining a rescaled peculiar--velocity field $\hat\bu = \bu / a \dot{b}$,
Eq.~(\ref{dust}b) becomes, after inserting Eq.~(\ref{parallelism}),
\begin{equation}
  \label{dustZA}
  \frac{d \hat\bu}{d b} = \b0 \;, \qquad  
  \left( \frac{d}{db} := \frac{\partial}{\partial b} + 
  \hat \bu \cdot \nabla_\bq \right) \;.
\end{equation}
The evolution according to this equation leads (kinematically) to the
formation of singularities, where $\varrho \rightarrow +\infty$, and
the velocity field becomes multi--valued. Following the idea of the
`adhesion model', the corrections to `dust' are assumed
to be negligible almost everywhere, i.e., except at the localized
singularities. The corrections are then estimated with the 
`dust' solution (i.e., with Zel'dovich's approximation):
\begin{list}{--}{\leftmargin=0pt \itemindent=5mm \itemsep=5mm}
\item {\bf The `EJN model'}: 
  inserting the parallelism condition~(\ref{parallelism}) into
  Eq.~(\ref{EJN}b), and 
  performing the same change of variables as before, we obtain, instead
  of Eq.~(\ref{dustZA}), the following corrected equation:
\begin{subequations}
  \label{burgers2} 
  \begin{equation}
  \frac{d \hat\bu}{d b} = {\mu} {\Delta}_\bq \hat\bu \;, 
  \qquad (\nabla_\bq \times \hat\bu = \b0) \;,
\end{equation}
with the GM or gravitational multi--stream coefficient (Buchert \&
Dom\'\i nguez 1998, Buchert et al. 1999)
\begin{equation}
  \label{GMcoeff}
  \mu := \frac{F}{a^2 {\dot b}} \lambda^2 (\varrho) > 0 \;,
\end{equation}
\end{subequations}
formally playing the role of a ``viscosity''.

\item {\bf The `SSE model'}: the parallelism condition allows us to
  express $\bw$ and $\varrho$ (via Eq.~(\ref{sse}c)) in terms of
  $\bu$. The correction term is hypothesized to be relevant only near
  places where Zel'dovich's approximation yields singularities, where
  $|\nabla \bu| \to \infty$, so that it can be approximated by
  neglecting subdominant powers of $\nabla \bu$:
  \begin{equation}
    \label{sseparallel}
    \frac{1}{\varrho} \left[ F_i - \frac{1}{a} \Pi_{ij,j} \right] 
    \approx \mbox{} - \frac{B L^2}{a (\nabla_\bq \cdot \bu)} 
    \partial_j [ (\nabla_\bq \cdot \bu) (\partial_k u_i) (\partial_k u_j) ] \; . 
    \end{equation}

    To simplify this expression further, one recalls that, within
    Zel'dovich's approximation, density singularities arise
    generically due to a locally plane--parallel collapse and the
    velocity gradient looks locally like $\partial_i u_j \approx
    (\nabla_\bq \cdot \bu) n_i n_j$, where the unit vector ${\bf n}$
    defines the plane of local collapse. 
 In order to compute the correction term, we assume that the
 spatial dependence of ${\bf n}$ is negligible compared to the (large)
 gradient of the fields along ${\bf n}$, i.e., as if it were a globally
 plane--parallel collapse, and we obtain:
    \begin{displaymath}
      \frac{1}{\varrho} \left[ F_i - \frac{1}{a} \Pi_{ij,j} \right] 
      \approx \mbox{} - \frac{3 B L^2}{a} (\nabla_\bq \cdot \bu) \Delta_\bq u_i \;\;.
    \end{displaymath}
    Therefore, Eq.~(\ref{sse}b) simplifies, after the change of variables,
    to the same Eq.~(\ref{burgers2}a) with a different GM coefficient:
    \begin{equation}
      \label{GM_SSE}
      \mu := \frac{3 B L^2}{b \varrho_H} \varrho > 0 \;.
    \end{equation}
    The dominant contribution in Eq.~(\ref{sseparallel}) is the one by
    velocity dispersion. It can be easily checked that the correction
    to the mean field gives, under the same assumptions, a subdominant but
    also positive contribution to the GM coefficient.

\end{list}

Eq.~(\ref{burgers2}) resembles the well--known key equation of the
original `adhesion approximation' (Gurbatov et al. 1989), to
which it reduces when $\mu$=constant $\rightarrow 0^+$. For constant
$\mu$ it can be solved analytically for curl--free flows (Hopf--Cole
transformation of the 3D Burgers' equation); this solution shows that the
singularities predicted by  Zel'dovich's approximation,
Eq.~(\ref{dustZA}), are indeed regularized by the $\mu$--term (see,
e.g., Vergassola et al.~1994); for an interesting application of Burgers' equation and further insight
see also the model of Jones (1996) for a two-component  system.
In the more general cases of a density--dependent GM coefficient, no
analytical solution was found. Nevertheless, the application of
boundary--layer theory (Dom\'\i nguez 2000) shows that density
singularities are still regularized (provided $p(\varrho \rightarrow
+\infty) \sim \varrho^\gamma$ with $\gamma>1$ in the case of the `EJN
model', so that pressure can oppose gravity successfully).  The
solutions for the velocity and density fields are qualitatively
identical to those of the original `adhesion model', developing into a
shock structure of infinite density in the limit $\mu \rightarrow
0^+$.

\smallskip

This derivation of the `adhesion approximation' offers insight into
the physics involved. The $\mu$--term, which was phenomenologically
motivated by analogy with viscosity, is not related to a truly
dissipative process.  
From a mathematical point of view, the starting point (`EJN
model'~(\ref{EJN}) or `SSE model'~(\ref{sse})), the approximations,
and the final Eq.~(\ref{burgers2}) are formally time--reversible,
i.e., invariant under the transformation $t \rightarrow -t$, $\bu
\rightarrow -\bu$. 
The correction to `dust' can transform mean kinetic energy ($\propto
\varrho |\bu|^2$) into internal kinetic energy ($\propto \Pi_{ii}$) and
conversely (also into internal gravitational potential energy via
$F_{i}$, but, as we have shown, the `SSE model' predicts this
contribution to be subdominant near density singularities in the
present approximation). The tendency to compression in collapsing
regions favors the correction to behave as a drain of mean kinetic
energy, mimicking viscous dissipation; it is easy to check that the
correction supplies energy in expanding regions 
(see \S\ref{sec:regular}), whose
expansion is thus accelerated.  This mechanism seems to be universal:
this is the reason that ``adhesiveness'' arises in the two different
models we study, `EJN' and `SSE', and that it can be modelled in a
qualitatively correct manner by the simplified `adhesion model' $\mu$ =
constant. The GM coefficient depends on time and on density, and this
leads to differences between models concerning the inner structure of
the regularized density singularities. A
density independent GM coefficient (as in the original `adhesion
approximation') is obtained with the imposed relationship $p \propto
\varrho^2$ corresponding to the naive application of the virial
condition, while the `EJN' and the `SSE' models coincide when $p
\propto \varrho^3$. 

\subsection{Relaxing key assumptions}
Summarizing, to derive Eq.~(\ref{burgers2}) one makes two important
assumptions: 
\begin{enumerate}
\item The mean peculiar--gravitational field $\bw$ for the evolution of
  `dust' is modelled by a first--order Lagrangian perturbation
  solution which includes parallelism of peculiar--velocity and
  \mbox{--acceleration} (Zel'dovich's approximation).
\item The so--approximated evolution of `dust' is used to estimate the
  correction to `dust', which for consistency must be assumed to be
  negligible everywhere except where singularities would develop. (For
  the `SSE model', one makes explicit use of the additional constraint
  of a locally plane--parallel singularity, which is nevertheless a
  consequence of Zeldovich's approximation and hence not expected to
  introduce new features.)
\end{enumerate}
Relaxing (1) means using a better approximation to the exact
evolution of `dust'. Zel'dovich's approximation, originally designed as an
extrapolation of the standard linear theory of gravitational instability into
the weakly non--linear regime, features ``slaving'' of peculiar--velocity 
and \mbox{--acceleration}. This simple, but physically
well--motived condition, if assumed initially, is
maintained up to second order in Lagrangian perturbation theory (see:
Buchert 1989, 1992 for the general first--order solution, 
Buchert \& Ehlers 1993 for explicit discussions within the general
second--order solution, and Buchert 1994 for the corresponding third--order solution).
Although corrections to  Zel'dovich's approximation
in general break the parallelism of $\bw$ and $\bu$, the departures are not
expected to be large in the perturbative regime, given that the 
`Zel'dovich approximation' performs well before shell--crossing. 
Nevertheless, using better approximations for the evolution of `dust' will not 
prevent singularities from appearing, because this is a general property of the 
`dust model' (see \S\ref{sec:nonpert}).

While relaxing (1) implies construction of sophisticated non--perturbative
approximations (since only then can we expect to substantially improve on the
performance of Lagrangian perturbation schemes, optimized to match N--body
simulations for any kind of dark matter),  
assumption (2) appears more like an unnecessary simplification. It is crucial
for the derivation of equations similar to the 
key equation of the standard `adhesion approximation', but 
it must be relaxed if one intends to improve over the `adhesion model'
in order to describe the subsequent dynamical evolution inside the
collapsing structures.
Assumption (2) may be rephrased by saying that the
evolution is dominated by convection, $({\hat \bu}\cdot \nabla_\bq)
{\hat \bu}$, over ``viscosity'', $\mu \Delta_\bq {\hat \bu}$ --- in
the hydrodynamic jargon, this is the limit 'Reynolds number
$\rightarrow +\infty$'.  Relaxing this assumption, i.e.~finite
Reynolds number, means taking account of the back--reaction of the
correction to `dust' on the trajectories of fluid elements away from
singularities too. In general, this will also imply corrections to the
parallelism of $\bw$ and $\bu$ as well as those due to the exact `dust'
evolution (see, e.g., Eq.~(\ref{EJNparallelism})).

Menci (2002) has studied the correction to the original
`adhesion approximation' when assumption (1) is relaxed. The starting
point is the set of Eqs.~(\ref{momenteq}) with the correction to
`dust' already modelled as $\mu {\Delta}_\bq \bu$, $\mu=$ constant
$\to 0$. The correction to parallelism, $\bw - F \bu$, is estimated
perturbatively in the limit of small velocities or times (the initial
condition is taken to satisfy parallelism) --- the procedure employed
by Menci {\em is not} an expansion in the ``viscosity'' $\mu$,
although it might formally appear so. The results agree somewhat
better with N--body simulations concerning the details of the forming
structures at small scales.

The solution to the `adhesion model', Eq.~(\ref{burgers2}a) with
constant $\mu$, cannot be computed as a regular expansion in $\mu$, because the naive
zeroth--order term, given by Eq.~(\ref{dustZA}), is undefined after shell--crossing.
The exact analytical solution to this model demonstrates
that the limit $\mu \rightarrow 0$ yields a {\it weak solution} of
Eq.~(\ref{dustZA}), exhibiting discontinuities in $\bu$. In the pedagogical review 
by Frisch \& Bec (2001) it is illustrated how much the Lagrangian solution of 
a 'multi--streamed dust model' differs from the Lagrangian solution to
the `adhesion model' after shell-crossing. In particular, it is
explained how the emergence of multi--streaming cannot be captured by a
Taylor expansion in the time variable of the solution to
Eq.~(\ref{dustZA}). As a matter of fact, the corresponding
perturbative expansion of the solution to the Vlasov--Poisson system
is identical to the one of the `dust model' {\em when the initial condition
is exactly single--streamed} (Valageas 2001)\footnote{Bharadwaj (1996)
  extended this claim to the case of {\em multi--streamed initial conditions}
  in the simplified model of free particles. This conclusion is, however,
  false because of a mathematical mistake in his Eq.~(34).}. This
problem can be overridden if a small 
amount of multi--streaming is allowed in the initial conditions, e.g.~in
the form of velocity dispersion.

Ribeiro \& Peixoto de Faria (2005) also attempted to provide a physical
motivation for the `adhesion model': starting from the
velocity potential $\varphi$ (with $\bu = \nabla_\bq \varphi / a$), it
is assumed that the complex variable $\psi=\sqrt{\varrho}
\exp{(i \varphi/\nu)}$ obeys a Schr\"odinger equation, where $\nu$ is
a constant with the appropriate dimension. From this assumption, it
is found that $\bu$ satisfies Eq.~(\ref{momenteq}b) with
a correction to `dust' that can be written as a functional of the density.
It is not obvious that with this correction term one can recover the `adhesion
model': it cannot be brought into a simple form $\propto \Delta_\bq
\bu$ after using the parallelism approximation, and the reasoning
offered by the authors is wrong from the outset because their
Eq.~(31) is algebraically false.

\smallskip

In the following sections we study the `EJN' and the `SSE' models beyond
assumptions (1, 2). 
First, we consider the standard Eulerian and Lagrangian perturbative
expansions, both closely related to the expansion in time just mentioned.
One of the key points in the derivation of the `adhesion
model' is the parallelism relation between $\bw$ and $\bu$,
Eq.~(\ref{parallelism}). This relation holds at the (Eulerian as
well as Lagrangian) linear order of the `dust model'.
Application of the perturbative techniques to the `EJN' and `SSE' models
will show how this relation is modified by the correction to `dust'.
Later we consider non--perturbative approaches: this is a
mathematically difficult task, and we are only able to collect some
results and provide some hints for future work.

\section{The Eulerian perturbative expansion}
\label{sec:euler}

The expansion parameter of the Eulerian perturbative expansion is the
amplitude of the departures of the initial conditions from homogeneity.
Formally, one writes:
\begin{equation}
  \label{eulerexpansion}
  \varrho = \sum_{n=0}^\infty \epsilon^n \varrho^{(n)} \;, \qquad
  \bu = \sum_{n=0}^\infty \epsilon^n \bu^{(n)} \;,
\end{equation}
where $\epsilon \ll 1$ is an expansion parameter, which can be taken to be
the variance of the initial density fluctuations, $\epsilon =
\sigma(t_0)$, with (note that $\sigma$ is finite because the
  density field has been smoothed over the scale $L$,
  Eq.~(\ref{defs}a))
\begin{equation}
  \sigma^2 (t) := \frac{\langle [\varrho(\b0, t) - 
  \varrho_H(t)]^2 \rangle}{\varrho^2_H (t)} \;.
\end{equation}
Here, $\langle \cdots \rangle$ denotes ensemble average over the
initial conditions.
Inserting these expansions into Eqs.~(\ref{momenteq}) and equating
powers of $\epsilon$, one obtains a set of linear equations for
$\varrho^{(n)}(\bq,t)$, $\bu^{(n)}(\bq,t)$. The zeroth order is the homogeneous
background, $\varrho^{(0)} (\bq, t) = \varrho_H (t)$, $\bu^{(0)} (\bq,
t) = \b0$. Due to the gravitational instability, the fields
$\varrho^{(n)}$, $\bu^{(n)}$ ($n \geq 1$) eventually
grow in time and the expansions~(\ref{eulerexpansion}) (likely to be understood
as {\em asymptotic} expansions) cease to be a useful representation of
the solution after some time $T$, which can be roughly estimated by the
condition $\sigma(T)=1$.

\subsection{The linear regime, ${\cal O}(\epsilon)$}
\label{sec:eulerlin}

To this order, one obtains the following equations:
\begin{subequations}
  \label{eulerlinear}
  \begin{equation}
    \frac{\partial \varrho^{(1)}}{\partial t} + 3 H \varrho^{(1)} + 
    \frac{1}{a} \varrho_H \nabla_\bq \cdot {\bu}^{(1)} = 0 \;,
  \end{equation}
  \begin{equation}
    \frac{\partial {\bu}^{(1)}}{\partial t} + H \bu^{(1)} = {\bw}^{(1)} + 
    \frac{1}{\varrho_H} \left( \bF^{(1)} - 
    \frac{1}{a} \nabla_\bq \cdot \bPi^{(1)} \right) \;,
  \end{equation}
  \begin{equation}
    \nabla_\bq \cdot {\bw}^{(1)} = - 4 \pi G a (\varrho^{(1)} - \varrho_H) \;, 
  \end{equation}
  \begin{equation}
    \nabla_\bq \times {\bw}^{(1)} = \b0 \;. 
  \end{equation}
\end{subequations}

For the `dust' and the `SSE' models, $\bF^{(1)} = \b0$ and
$\bPi^{(1)}=0$, and they do not differ at this order. 
One can show that the vorticity, $\boldsymbol{\omega} = \frac{1}{a}\nabla \times \bu$, decays
with time, and that the peculiar--velocity becomes parallel to the 
peculiar--acceleration asymptotically in time,
\begin{equation}
  \label{linearparallelism}
  \bw^{(1)} = F(t) \bu^{(1)} \;.
\end{equation}
For the `EJN model', $F_i^{(1)} = 0$, but:
\begin{equation}
  \label{pilinear}
  \Pi_{ij}^{(1)} = p'(\varrho_H) \, \varrho^{(1)} \,  \delta_{ij} \;.
\end{equation}
Then, Eqs.~(\ref{eulerlinear}) can be simplified to a single equation for
the density contrast $\delta := \varrho/\varrho_H - 1$:
\begin{equation}
  \label{deltalinear}
  \frac{\partial^2 \delta^{(1)}}{\partial t^2} + 2 H \frac{\partial \delta^{(1)}}{\partial t} - 
  4 \pi G \varrho_H \delta^{(1)} = 
  4 \pi G \varrho_H \frac{\lambda^2 (\varrho_H)}{a^2} \Delta_\bq \delta^{(1)} \;,
\end{equation}
where we have used the length defined in Eq.~(\ref{lambda}). Notice that
in the absence of cosmological expansion, the time--dependence of
$\lambda$ induced by $\varrho_H$ drops and it reduces to the usual
definition of Jeans' length (up to a factor $2 \pi$, e.g.~Peebles~1980).
Eq.~(\ref{deltalinear}) can be solved analytically for particular
choices of the cosmological background and the dependence
$p(\varrho)$ (Haubold et al. 1991). Qualitatively, one finds that
the pressure term is irrelevant for inhomogeneities on scales $\gg
\lambda$, which grow according to the `dust model', while those on scales
$\ll \lambda$ are damped by pressure and may exhibit sound--like oscillations.

The `EJN model' provides a correction to the parallelism relation
already at the linear order. Following Buchert et al. (1999)
\footnote{In that work, only the particular case of a polytropic
  dependence, $p(\varrho) \propto \varrho^\gamma$, in an Einstein--de
  Sitter cosmological background was considered, but it can be checked
  that Eq.~(\ref{EJNparallelism}) holds generally. Also in that work,
  a minor mistake was made and the coefficient ${\cal C}(t)$ quoted there 
  is missing an extra time--dependent factor when
  $\gamma>13/6$.}, the solution to Eq.~(\ref{deltalinear}) and the
departure from parallelism can be computed as a series in powers of
$\lambda$ (unlike in the fully non--linear problem, the limit $\lambda
\rightarrow 0$ in Eq.~(\ref{deltalinear}) is regular).  Asymptotically
in time, we may write (${\cal C}(t)$ is a dimensionless
time--dependent coefficient):
\begin{eqnarray}
  \label{EJNparallelism}
  \bw^{(1)} & = & F (t) \left[ \bu^{(1)} -  
    \frac{{\cal C}(t)}{a^2} \lambda^2(\varrho_H) \Delta_\bq \bu^{(1)} 
    + {\cal O}(\lambda^4) \right] \nonumber \\
    & = & F(t) \bu^{(1)} + {\cal C}(t) \frac{p'(\varrho_H)}{a \varrho_H} 
    \nabla_\bq \varrho^{(1)} + {\cal O}(\lambda^4) \;,
\end{eqnarray}
where the last equality follows by using
Eqs.~(\ref{eulerlinear}c,\ref{lambda}). The correction is proportional to
the extra acceleration due to the pressure gradient. Since ${\cal C}(t) \neq 1$
in general, the peculiar--velocity is not parallel to the total peculiar--acceleration
(gravity+pressure).

\subsection{The non--linear orders, ${\cal O}(\epsilon^n)$, $n>1$}
\label{sec:eulernonlin}

The Eulerian perturbative expansion beyond the linear order has been
intensively studied for the `dust model' in recent years (see,
e.g., Bernardeau et al. 2002 and references therein). The predictions of the 
`SSE model' differ from those of the `dust model' at non--linear orders. 
An important difference concerns the vorticity. With full generality 
(i.e., to all orders of the perturbative expansion), one can easily show from
Eq.~(\ref{momenteq}b) 
by application of the vector identity $(\bu \cdot \nabla_\bq) \bu = (1/2) \nabla_\bq
|\bu|^2 - \bu \times (\nabla_\bq \times \bu)$ that (here, a comma denotes partial
derivative with respect to comoving Eulerian coordinates)
\begin{align}
  \label{vort}
  \frac{\partial \omega_i}{\partial t} + 2 H \omega_i + & \frac{1}{a} 
  \left( u_j \omega_{i,j} - u_{i,j}\omega_j +  u_{j,j} \omega_i \right) = \nonumber \\
  & \frac{1}{a} \varepsilon_{ijk} \, \partial_j \left[ \frac{1}{\varrho} 
  (F_k - \frac{1}{a} \Pi_{km,m}) \right] . 
\end{align}
For the `dust model' (vanishing r.h.s.), this equation
can be integrated exactly along the flow lines, Eq.~(\ref{trajectory})
(e.g., Serrin 1959, Buchert 1992, Eq.~(26){\it ff}):
\begin{equation}
  \label{eq:vort_dust}
\omega_i (\bX ,t) = \left(\frac{a_0}{a}\right)^2 \;
\frac{\omega_j (\bX,t_0) 
  \left[ \delta_{ij} + \frac{\partial P_i (\bX ,t)}{\partial X_j} \right]}
{\left|\det \left( \delta_{ij} + \frac{\partial P_i (\bX ,t)}{\partial X_j} \right) \right|} \; ,
\end{equation}
We see that only the corrections to `dust' can be a source of vorticity: 
if the initial vorticity vanishes (as is usually
assumed in cosmological models of structure formation by virtue of the
decay of the linear vorticity), then the evolution according to the
`dust model' predicts that the vorticity remains zero (Kelvin's
circulation theorem). This remains true for the `EJN model', because
the pressure is represented as a function of density only and the
r.h.s. of Eq.~(\ref{vort}) vanishes. The `SSE model',
however, does incorporate the non--linear generation of vorticity: when
Eq.~(\ref{vort}) is evaluated to order ${\cal O}(\epsilon^2)$ with the
expansions~(\ref{eulerexpansion}), one obtains:
\begin{equation}
  \frac{\partial \omega_i^{(2)}}{\partial t} + 2H \omega_i^{(2)} = 
  BL^2 \varepsilon_{ijk} \left[ \varrho_{,jm}^{(1)} w_{k,m}^{(1)} - 
    \frac{1}{a} (u_{m,l}^{(1)} u_{k,l}^{(1)})_{,jm} \right] \;.
\end{equation}
The source terms of vorticity are the tidal and shear stretching and can
be evaluated as a functional solely of the linear density field
$\varrho^{(1)}$ by resorting to Eqs.~(\ref{eulerlinear}c,
\ref{linearparallelism}). A detailed study of the growth of the
ensemble average $\langle
\omega^2 \rangle$ and its dependence on the initial conditions can be
found in (Dom\'\i nguez 2002). As soon as the velocity field has a
non--vanishing vortical component, parallelism with the gravitational
peculiar--acceleration is rigorously ruled out. (But this does not mean that
deviations from parallelism must be necessarily large or important).
Vorticity acts against collapse (see \S\ref{sec:nonpert}) and is
likely to play a role in shaping the internal structure of collapsed
regions, which in the `adhesion model' 
are structureless.

\section{The Lagrangian perturbative expansion}
\label{sec:lagrange}

To perform the Lagrangian perturbative expansion, we have to transform the
equations with respect to Lagrangian coordinates and follow the field
quantities along comoving trajectories of the fluid elements:
\begin{equation}
\bq = \bX + \bP(\bX,t) , \qquad \bu=a {\dot\bP} , \qquad \bP(\bX,t_\init)=\b0 \;,
\label{trajectory}
\end{equation}
with the Lagrangian time--derivative operator  $\; \dot{ } :=
\frac{\partial}{\partial t}\vert_\bX =\frac{\partial}{\partial t}\vert_\bq + \frac{u_i}{a}
\frac{\partial}{\partial q_{i}}$.
(Note that $\bX + \bP$ are integral curves of the {\em mean}
peculiar--velocity, scaled by $a(t)$). $\bX$ are Lagrangian
coordinates and $\bP(\bX,t)$ is the displacement of the fluid volume
elements away from their initial positions; this is the expansion parameter in
the Lagrangian perturbative approach. This
method has been applied successfully to study the `dust model', 
Zel'dovich's approximation being a subcase of the first--order
solution, as well as higher--order solutions. 
It has been recently applied to the `EJN model' too (see the
references below).  Concerning the `SSE model', there is no difference
with the `dust model' to linear order.

In view of later considerations,
for the application of the Lagrangian methods we take an
unconventional route and first rewrite Eqs.~(\ref{momenteq}) as
follows.
We obtain an evolution equation for the peculiar--gravitational field
$\bw$ by eliminating the density in Eq.~(\ref{momenteq}a) with the
help of Eq.~(\ref{momenteq}c).
By formally integrating the divergence one gets (Buchert 1989):
\begin{subequations}
\label{new}
\begin{gather}
\dot{\bw} + 2H\bw - 4\pi G \varrho_{H} \bu = \vec{\cal R}\;, \\
\vec{\cal R}: = \frac{1}{a} [(\bu \cdot {\nabla}_\bq) \bw -
\bu({\nabla}_\bq \cdot \bw) + 
{\nabla}_\bq \times \bT] \;. \nonumber
\end{gather}
The vector $\bT$ is essentially fixed by the other equation satisfied by $\bw$,
Eq.~(\ref{momenteq}d): acting with $\nabla_\bq\times$ on
Eq.~(\ref{new}a) and subjecting $\bT$ to the Coulomb gauge condition,
${\nabla}_\bq \cdot \bT = 0$, one finds
\begin{eqnarray}
\Delta_\bq \bT & = & 4\pi Ga {\nabla}_\bq \times (\varrho \bu)
= 4\pi Ga [ \varrho {\nabla}_\bq \times \bu + {\nabla}_\bq\varrho \times \bu ] 
\nonumber \\
& = & (4\pi Ga \varrho_H - {\nabla}_\bq \cdot \bw) \, {\nabla}_\bq \times \bu + 
\bu\times{\Delta}_\bq \bw \;.
\end{eqnarray}
That is, $\bT$ is the vector potential (up to an unimportant factor)
of the peculiar--current density ${\bf j}: =\varrho \bu$ (i.e., $4 \pi
G a {\bf j}=-\nabla_\bq \times \bT-\nabla_\bq \,\frac{1}{a}\partial_t (a\phi)$, with $\phi$
denoting the scalar peculiar--gravitational potential). 
(If $\bw = :-\frac{1}{a}\nabla_\bq \phi$ is initially
irrotational, Eqs.~(\ref{new}a,b) propagate this condition in time.)
Finally, Eq.~(\ref{momenteq}b) gives the evolution of $\bu$:
\begin{equation}
  \dot{\bu} + H \bu = \bw + \frac{1}{\varrho} 
  \left( \bF - \frac{1}{a} \nabla_\bq \cdot \bPi \right)\;.
\end{equation}
\end{subequations}
Thus, we have replaced Eqs.~(\ref{momenteq}) for $\varrho$, $\bu$, and
$\bw$ by the new set of Eqs.~(\ref{new}) for $\bu$ and $\bw$.  They
hold independently of what the correction to `dust' looks like. These
equations still retain the nonlocal nature imposed by gravity through
the vector potential $\bT$ given by Eq.~(\ref{new}b); we obtain
``local'' evolution equations only if the source of this equation
vanishes. Physically this means that the peculiar--current density has
to be irrotational, and, for irrotational flows, that the mean flow
follows the gradient of the density field\footnote{If such conditions
  are imposed on the problem, $\bT$ becomes a harmonic vector function
  which can be set to zero for periodic boundary conditions of the
  cosmological fields on some large scale (since then the only
  harmonic functions are spatially constant and can be set to zero due
  to the invariance of the basic equations with respect to spatially
  constant translations).}.  In the following, we will not be further
concerned with the residual vector field $\vec{\cal R}$.  We just
acknowledge that it is a functional of the dynamical fields $\bu$ and
$\bw$ so that the system of equations (\ref{new}a-c) is closed for
given $\bF$ and $\bPi$.

\subsection{The single--streamed case of `dust matter'\label{sec:dustLang}}

For the `dust model' the system of equations
(\ref{new}a-c) has been studied thoroughly in (Buchert
1989). There, the currently used notion of the {\em weakly non--linear
regime} has been defined as ``Lagrangian linearization''.
The comoving trajectory field $\bP$ of Zel'dovich's approximation 
(Zel'dovich 1970, 1973) can be understood as a subclass of the
linear solutions in Lagrangian space (Buchert 1989, 1992). For an Einstein--de Sitter cosmology
with initial peculiar--velocity $\bu(\bX, t_\init)$, the subclass 
corresponding to Zel'dovich's model is 
$\bP(\bX, t) = \frac{3}{2} [b(t)-b(t_\init)] \bu(\bX, t_\init) \, t_\init$, where
$b(t)=a(t)$ here.
For all background cosmologies including ``curvature'' and a cosmological 
constant Zel'dovich's approximation is given in (Bildhauer et al. 1992; for
the flat models with cosmological constant, see also the supplement by 
Chernin et al. 2003).

This well--known result can be easily obtained from the general
system of equations (\ref{new}a-c) by the following reasoning:
in the Lagrangian picture the Eulerian dynamical variables
$\bu(\bq,t)$ and $\bw(\bq,t)$ are replaced by the single dynamical
variable $\bP(\bX,t)$.  Lagrangian linearization means that the
equations have to be linearized with respect to $\bP$. 
Eqs.~(\ref{trajectory}) and (\ref{new}c) (without corrections to `dust')
express the fields $\bu$
and $\bw$ as linear functions of $\bP$. It follows that Eqs.~(\ref{new}a,b)
have to be linearized in the fields $\bu$ and $\bw$.
For simplicity we restrict the initial conditions to
irrotational flow, so that the velocity remains irrotational through
the `dust' evolution, Eq.~(\ref{eq:vort_dust}).
Then, Eq.~(\ref{new}b) implies, upon linearization, $\bT=\b0$.
From (\ref{new}a) we can therefore immediately drop the manifestly
non--linear terms in the residual vector field $\vec{\cal R}$. 
Introducing the longitudinal and transversal parts of $\bu$ and $\bw$ with respect
to Lagrangian coordinates, e.g., ${\nabla}_{\bf 0}\cdot\bw^T = 0$, 
${\nabla}_{\bf 0}\times\bw^L = {\bf 0}$,
where ${\nabla}_{\bf 0}$ denotes the nabla operator 
with respect to Lagrangian coordinates, the
remaining equations for `dust' read: 
\begin{eqnarray}
  \label{dustLangLinear}
  &{\dot\bw^L} + 2H {\bw^L} - 4\pi G \varrho_H \bu^L = {\bf 0} \;\;,\;\; 
  {\dot\bu^L} + H {\bu^L} - \bw^L = {\bf 0} \;, \nonumber\\
  &\bu^T = \bw^T = \b0\;.
\end{eqnarray}
These equations are satisfied if the inhomogeneous displacement field 
$\bP$, split into its longitudinal and transversal parts, 
obeys: 
\begin{gather}
  \label{pepsidust}
  {\ddot\bP^L} + 2 H {\dot\bP^L} - 4\pi G \varrho_H \bP^L = {\bf 0} \;, \\
  \bP^T =  {\bf 0} \;, \nonumber
\end{gather}
which can be easily demonstrated by reinserting the definitions of 
the fields $\bu$ and $\bw$ in favor of $\bP= \bP^L + \bP^T$. If we restrict the general
solution of Eq.~(\ref{pepsidust}) to the initial data $\bw(\bX,t_\init) = 
({\it const.}/t_\init) \, \bu(\bX,t_\init) \;\;(const. = 1$ for an Einstein--de Sitter
cosmology),  we obtain Zel'dovich's approximation. The rigorous calculation, 
starting from the fully 
transformed system of equations, may be found in (Buchert 1989, 1992).

\subsection{The multi--streamed case modelled by dynamical stresses}

In the regime when the `dust model' predicts multi--streaming, the
corrections to `dust' become relevant. 
We look for the Lagrangian linear form by also dropping the 
residual vector field $\vec{\cal R}$ (since initial irrotationality is
preserved to linear order by the evolution in both the `EJN' and the
`SSE model').
The `SSE'
correction to `dust' is also non--linear, Eq.~(\ref{sse}b), so that this
model does not differ from the `dust model' to linear order.
For the `EJN model', Eq.~(\ref{EJN}b), however,
we have to linearize additionally the  
(Eulerian) Laplacian together with the GM~coefficient 
along the comoving trajectory field. Using 
transformation tools developed by Adler \& Buchert (1999, Appendix A) 
we find (for isotropic $\bPi$):
\begin{equation}
\frac{\lambda^2 (\varrho)}{a^2} {\Delta}_\bq\bw = 
\frac{\lambda^2 (\varrho_H)}{a^2} {\Delta}_\b0 \bw + 
{\cal O}(\bP^2) \;.
\label{deltatrafo} 
\end{equation}
The complete set of Lagrangian linear equations may then be cast into
a form that reproduces the `dust' result~(\ref{dustLangLinear}) on their left--hand--sides:
\begin{gather}
  \label{lagEJN}
  {\dot\bw^L} + 2H {\bw^L} - 4\pi G \varrho_H \bu^L = {\bf 0} \;, \nonumber \\
  {\dot\bu^L} + H {\bu^L} - \bw^L = \frac{1}{a^2}\lambda^2 (\varrho_H)
  {\Delta}_{\bf 0} \bw^L \;, \\
  \bu^T = \bw^T =\b0 \;. \nonumber
\end{gather}
Again, the rigorous way of deriving the equation obeyed by Lagrangian linear
displacements is the following: we first transform the basic equations into 
Lagrangian form, then
linearize the Lagrangian system with respect to $\bP$. 
This has been done in (Adler \& Buchert 1999) with the result that 
$\bP$ obeys:
\begin{eqnarray}
&{\ddot\bP^L} + 2 H {\dot\bP^L} - 4\pi G \varrho_H \bP^L = 
4 \pi G \varrho_H \frac{\lambda^2 (\varrho_H)}{a^2} {\Delta}_{\bf 0}\bP^L \, , \nonumber \\
&{\ddot\bP^T} + 2 H {\dot\bP^T} ={\bf 0}\;;\;\;\;{\rm here:}\;\;\;\bP^T ={\bf 0}\;\;.
\label{pepsi}
\end{eqnarray}
After some (in this case more involved) manipulations one can
find the equations for $\bu$ and $\bw$ above by inserting their
definitions in terms of $\bP$ into Eqs.~(\ref{pepsi}).
The differential operator acting on $\bP^L$ is formally identical to 
Eq.~(\ref{deltalinear}), and Eq.~(\ref{pepsi}) 
can thus
be solved by extrapolation of the known Eulerian linear solutions
(Adler \& Buchert 1999). 

Evidently, the adhesive term consists of a Lagrangian Laplacian and,
therefore, it involves, in addition to the convective non--linearities
hidden in the overdot, non--linearities in Eulerian space
when mapping it back 
using the inverse solution of the mapping $\bq = \bX + \bP (\bX,t)$.

In general, parallelism of peculiar--velocity and
\mbox{--acceleration} is violated. By computing the time derivative of
the difference $\bw - F(t) \bu$ with Eqs.~(\ref{pepsi}), one can show
that in general it does not decay to zero asymptotically for large
times. The correction is proportional to ${\Delta}_{\b0} \bu$ to
lowest order in $\lambda^2$. 
This correction is similar to the one found in
the Eulerian linear approximation, Eq.~(\ref{EJNparallelism}), but in
terms of the Lagrangian gradients.

\subsection{Beyond first order}

The Lagrangian perturbation scheme including pressure (Adler \& Buchert 1999)
has been pushed to second order (Morita \& Tatekawa 2001; Tatekawa et al. 2002)
and recently to third order (Tatekawa 2005).
Results on cross--correlation statistics with N--body
simulations (Tatekawa 2004a) and a comparison of the corresponding density 
fluctuations with results of hydrodynamical simulations (Tatekawa 2004b)
has demonstrated that multi--stream forces tend to be
underestimated by Lagrangian perturbative models.
In the next section we shall investigate why this happens.

\section{The non--perturbative regime}
\label{sec:nonpert}

\subsection{Exact evolution equations I: general case}
\label{sec:exactgeneral}

Following a systematic perturbative approach we could provide models
for adhesive gravitational clustering in the weakly non--linear regime.
This has allowed us also to relax the parallelism assumption 
(\ref{parallelism}) that was a necessary ingredient of the standard
`adhesion model' but, as we have shown, is not expected to hold 
in the multi--streamed regime. 

Finding an extension into the non--perturbative (both Eulerian and
Lagrangian) regimes is an involved mathematical task. Notwithstanding,
it should be attempted in view of the fact that even the Lagrangian
perturbation approach falls short of capturing the action of
multi--stream forces, since a genuine property of adhesive models in
the simplest cases involves an Eulerian Laplacian. Thus, we expect the
best approximations to be of hybrid Lagrangian/Eulerian type like the
standard `adhesion model',
meaning a non--linear equation both in Eulerian space (due to the
convective non--linearities) as well as in Lagrangian space (due to the
Eulerian gradients in the forcing).

An exact equation for the comoving deviations of the trajectory field,
$\bP(\bX,t)$ defined by Eq.~(\ref{trajectory}), can be obtained from
Eqs.~(\ref{new}): formal integration of Eq.~(\ref{new}a) yields
\begin{equation}
  \label{w_general}
 \bw (\bX, t) = 4 \pi G a \varrho_H \bP(\bX,t) + 
 \frac{1}{a^2} \int_{t_\init}^t \!\!\! d\tau \; a^2(\tau) \vec{\cal R}(\bX,\tau) ,
\end{equation}
with the integration constant chosen such that the Lagrangian
coordinates are defined by the initial datum for $\bw$,\footnote{For details on the
possibility and advantage of this choice, see Adler \& Buchert 1999, Appendix A.} 
$4 \pi G a_\init
\varrho_H (t_\init) \bP(\bX, t_\init) = \bw(\bX, t_\init)$.
Then, from Eq.~(\ref{new}c) one obtains the general result:
\begin{align}
  \label{P_general}
  \ddot{\bP} + 2 H \dot{\bP} - 4 \pi G \varrho_H \bP &
  - \frac{1}{a^2} \int_{t_\init}^t \!\!\! d\tau \; a^2(\tau) \vec{\cal R}(\bX,\tau) \nonumber \\
  & \mbox{} = \frac{1}{a \varrho} \left[ \bF - \frac{1}{a} \nabla_\bq \cdot \bPi \right] \; .
\end{align}
The left--hand--side of this equation reproduces the `dust model'. If
the correction to `dust' is expressed in terms of the fields
$\varrho$, $\bu$, $\bw$, then the equation can be written entirely in terms of
$\bP$ and $\vec{\cal R}$ by virtue of
Eqs.~(\ref{trajectory},\ref{w_general},\ref{rho_general}). Thus, the
difficulty in deriving a closed, local differential equation for the deviation field
$\bP$ lies in the term $\vec{\cal R}$, which is a nonlocal (in space and time) and a
non--linear functional of $\bP$ given by Eq.~(\ref{new}b).

As a further illustration of the general case we write down the corresponding equation
for the density. For this purpose we introduce the {\em contrast density field} 
$\Delta : = (\varrho_H - \varrho)/\varrho$, $-1 < \Delta < \infty$, which is
more adapted to the non--linear situation than the conventional density
contrast $\delta  = (\varrho - \varrho_H )/\varrho_H$ (for the `dust' case see: 
Buchert 1989, Eq.~(7){\it ff}; 1992 Eq.~(31){\it ff}; 1996 Eq.~(25){\it ff} 
(with a different sign convention for $\Delta$)).
The exact evolution equation for this variable can be obtained
from Eqs.~(\ref{momenteq}a,b):
\begin{multline}
  \label{beyondquasilinear2}
\ddot \Delta + 2 H \dot \Delta - 4\pi G \varrho_H \Delta = \\
(1 + \Delta) \left[ 2 II^{\rm pec} + \frac{1}{a} {\nabla}_{\bq}\cdot\left(
\frac{1}{\varrho}\bF - 
\frac{1}{\varrho a}{\nabla}_{\bq}\cdot\bPi \right)\right]\; ,
\end{multline}
with the second scalar invariant of the peculiar--velocity gradient,
\begin{displaymath}
II^{\rm pec}:= \frac{1}{2 a^2}[(u_{i,i})^2 - u_{i,j}u_{j,i}] = 
\frac{1}{2 a^2}{\nabla}_{\bq} \cdot (\bu 
{\nabla}_{\bq}\cdot \bu - \bu \cdot {\nabla}_{\bq} \bu ) \;\;.  
\end{displaymath}

\vspace{5pt}

For the purpose of evaluating the Eulerian spatial derivatives in the 
right--hand--side of the above general equations, we employ the following
transformation: 
\begin{equation}
  \label{derivative_general}
  \frac{\partial}{\partial q_i} = (J^{-1})_{ji} \, \frac{\partial}{\partial X_j}\;,
\end{equation}
where $J^{-1}$ is the inverse to the following matrix:
\begin{equation}
  \label{Jmatrix_general}
  J_{ij}(\bX,t) := \delta_{ij} +  \frac{\partial P_i(\bX,t)}{\partial X_j} \;.
\end{equation}
In the rest of this section we assume always that the map $\bX \to \bq$
is invertible, i.e., no shell--crossing, so that $\det J_{ij} >0$.  
Mass conservation, Eq.~(\ref{momenteq}a),
relates the density field to $\bP$ via the matrix $J_{ij}$\footnote{Of course, 
$\varrho$ solves Eq.~(\ref{beyondquasilinear2}), if $\bP$ solves Eq.~(\ref{P_general}).}:
\begin{eqnarray}
  \label{rho_general}
  \varrho(\bX, t) & = & \left[\frac{a_\init}{a}\right]^3 \varrho (\bX, t_\init) \,
  \frac{\det J_{ij} (\bX,t_\init)}{\det J_{ij} (\bX,t)} \nonumber \\
  & \approx & \varrho_H(t) / \det J_{ij} (\bX,t) ,
\end{eqnarray}
where we take initial quasi--homogeneity for simplicity ($\varrho
(\bX,t_\init) \approx \varrho_H (t_\init)$, $J_{ij}(\bX,t_\init) \approx \delta_{ij}$).

\subsection{Exact evolution equations II: plane symmetry}
\label{sec:exactplane}

The problem is simplified if one considers particular
geometric settings with high symmetry. Although this leads to
not fully realistic models, it can provide hints of the mathematical
and physical properties of the models in the non--perturbative regime.
Plane--symmetric models offer an
excellent test case for numerical simulations of multi--stream systems concerning a variety of 
properties including spatial scaling 
(e.g. Doroshkevich et al. 1980; Gouda \& Nakamura 1988, 1989; Yano \& Gouda 1998; Fanelli \&
Aurell 2002; Aurell et al. 2003; Yano et al. 2003; the assumption of 
spherical symmetry plays a complementary role, e.g. Padmanabhan 1996.)
In such configurations, the fields vary spatially along a single direction, and so
render the Lagrangian approach
particularly suitable for an extension into the non--perturbative
regime. Plane--symmetric models simplify the problem enormously, since 
the residual term $\vec{\cal R}$ vanishes, and the `dust' evolution remains in the
Lagrangian linear regime at all times (implying in
particular that the parallelism relation~(\ref{parallelism}) --- stating a 
spatially constant factor of proportionality --- holds
asymptotically at large times).

In a plane--symmetric configuration, the fields vary only along a
single direction, say $i=1$. 
Eqs.~(\ref{w_general},\ref{P_general},\ref{derivative_general},\ref{rho_general})
then are simplified to (we hereafter drop the index $i=1$):
\begin{eqnarray}
  \label{1drelations}
  \varrho(X, t) & = & \varrho_H(t) \, \left[ 1 + \frac{\partial P(X,t)}{\partial X} 
   \right]^{-1} \;,\nonumber \\
  \frac{\partial}{\partial q} & = & \left[ 1 + \frac{\partial P(X,t)}{\partial X} \right]^{-1} 
  \frac{\partial}{\partial X}\;, \nonumber \\
  w (X, t) & = & 
  4 \pi G a \varrho_H P(X,t) \;,\\
  \ddot{P} + 2 H \dot{P} - 4 \pi G \varrho_H P & = &
  \frac{1}{a \varrho} F - \frac{1}{a^2 \varrho_H} \frac{\partial \Pi}{\partial X}\;. \nonumber 
\end{eqnarray}
Therefore, depending on the specific model, we get different equations:
\begin{itemize}
\begin{subequations}
\label{1dP}
\item `Dust model':
  \begin{equation}
     \ddot{P} + 2 H \dot{P} - 4 \pi G \varrho_H P = 0 \;.
  \end{equation}
\item `EJN model': inserting Eqs.~(\ref{EJNcorrections})
  particularized to the plane--symmetric configuration, one obtains:
  \begin{equation}
     \ddot{P} + 2 H \dot{P} - 4 \pi G \varrho_H P = 
     - \frac{1}{a^2 \varrho_H} \frac{\partial}{\partial X} \,
    p \left( \frac{\varrho_H}{1+\partial P/\partial X} \right) \;.
  \end{equation}
\item `SSE model': using Eqs.~(\ref{SSEcorrections}), 
  \begin{align} 
     \ddot{P} & + 2 H \dot{P} - 4 \pi G \varrho_H P = 
     - B L^2 \frac{\partial}{\partial X} \left\{
      \frac{(\partial \dot{P}/\partial X)^2}{(1+\partial P/\partial X)^3} \right. \nonumber \\
      & \mbox{} + \left. 4 \pi G \varrho_H \left[ \frac{1}{2 (1+\partial P/\partial X)^2} - 
        \frac{1}{1+\partial P/\partial X} \right] \right\} \;.
  \end{align}
  \end{subequations}
\end{itemize}
Eqs.~(\ref{1dP}) are closed partial differential equations
for the function $P(X,t)$. 
Notice that the equation for the `dust model' is linear, so that Lagrangian
linearization already yields the exact solution.
The equation for $P(X,t)$ in the `EJN model'
becomes also a linear equation for the special
dependence $p \propto \varrho^{-1}$ (Chaplygin gas, App.~\ref{app:eqstate};
{\it cf.} Adler \& Buchert 1999).
This case is formally of interest because it provides a subcase in which the
Lagrangian linear solution exactly solves the plane--symmetric problem.  

Eqs.~(\ref{1dP}) should allow one to explore the role of the corrections
to `dust' beyond the assumptions employed in \S\ref{sec:adhesion}, in
particular, the back--reaction on the `dust' trajectories by the
correction terms. It should be noted that Eqs.~(\ref{1dP}) 
are of second order in time, in contrast to the generalized
`adhesion model', Eq.~(\ref{burgers2}a)
--- the relevance of this fact concerning
the formation of singularities is discussed in \S\ref{sec:regular}. In
this connection Eqs.~(\ref{1dP}b,c) have features of a
quasilinear hyperbolic (``wave'') equation, while Eq.~(\ref{burgers2}a) is a quasilinear parabolic
(``diffusion'') equation. 
G\"otz (1988) considered the `EJN model' with the
``isothermal'' relationship $p \propto \varrho$ in the absence of
background cosmological expansion ($a(t) \equiv 1$), 
and he was able to show that it can be mapped exactly to the Sine--Gordon
equation, which is known to admit soliton solutions.

Fanelli \& Aurell (2002) studied numerically the one--dimensional
Vlasov--Poisson system (see \S\ref{sec:coarsening}), in particular the
time--dependence of a collapsing isolated perturbation. Fanelli \&
Aurell tried to interpret the numerical findings in the framework of
the simplified `EJN model'~(\ref{burgers2}) with a polytropic
relationship for the pressure, $p(\varrho) \propto \varrho^\gamma$. A
simple scaling argument led them to the value $\gamma=7/9$ (the
original `adhesion model' corresponding to $\gamma=2$). This
conclusion is, however, questionable because boundary--layer theory
cannot be applied to Eqs.~(\ref{burgers2}) when $\gamma<1$. A more
general analysis (Dom\'\i nguez, unpublished) shows that a
shock can form in this case only when the initial velocity is smaller
than the maximum velocity, and this maximum velocity vanishes in the
limit $\mu \rightarrow 0$. That is, when $\gamma<1$, the adhesive
term is too weak to prevent singularities for arbitrary initial
velocities. Thus, it might be the case that Eqs.~(\ref{burgers2})
cannot describe even qualitatively the numerical experiment by
Fanelli \& Aurell.

\subsection{A new non--perturbative approximation}

As remarked, one of the difficulties with the exact
Eq.~(\ref{P_general}) is the nonlocal nature introduced by $\vec{\cal
  R}$. A possible approximation consists in setting
$\vec{\cal R}=\b0$, so that Eq.~(\ref{P_general}) simplifies to
\begin{equation}
  \label{P_approx}
  \ddot{\bP} + 2 H \dot{\bP} - 4 \pi G \varrho_H \bP 
  = \frac{1}{a \varrho} \left[ \bF - \frac{1}{a} \nabla_\bq \cdot \bPi \right] \; .
\end{equation}
As long as $\bF$ and $\bPi$ are modelled as local functions of
$\varrho$, $\bu$ and $\bw$ (as in the models we consider in this
work), this is a partial differential equation for $\bP$ by virtue of
Eqs.~(\ref{trajectory},\ref{w_general},\ref{rho_general}),
albeit a difficult one. 

One can view the approximation $\vec{\cal R}=\b0$ in the same spirit
as the parallelism assumption (\ref{parallelism}): one gets rid of the
problem of nonlocality by approximating the exact peculiar--gravitational field by 
a relationship which holds in the (both
Eulerian and Lagrangian) linear regime of the `dust model' as well as
exactly in the plane--symmetric configuration:
\begin{equation}
  \label{w_approx}
  \bw = 4 \pi G a \varrho_H \bP ,
\end{equation}
to be compared with $\bw = 4 \pi G a \varrho_H (b/\dot{b}) \dot{\bP}$
corresponding to Eq.~(\ref{parallelism}). 
The relation (\ref{w_approx}) replaces the ``slaving'' assumption of parallelism of
$\bw$ to the peculiar--velocity. It is more general and simpler, because it does not involve
specification of the function $b(t)$. 
Unlike the condition of parallelism, this approximation also retains features of the 
exact field equations for $\bw$: e.g., at any given instant of time, the approximated field
$\bw$ depends only on the displacement (i.e., the distribution of mass) at this time,
and not on its time--derivative (the peculiar--velocity field).
This renders the differential equation obeyed by $\bP$
second--order in time (see \S\ref{sec:regular} in this respect). 
A proper understanding of the extrapolation of the linear relationship between
$\bw$ and $\bP$ into the non--perturbative regime requires that the
residual term $\vec{\cal R}$ be analyzed, a task beyond the goals of
this work.

To get a glimpse of the mathematical difficulties posed by
Eq.~(\ref{P_approx}), we particularize it to the `EJN model' of $\bF$
and $\bPi$, Eqs.~(\ref{EJNcorrections}). The right--hand--side can be
written, in analogy with Eq.~(\ref{EJN}b), as:
\begin{equation}
\label{particularizing}
  \frac{1}{a \varrho} \left[ \bF - \frac{1}{a} \nabla_\bq \cdot \bPi \right] =
  \frac{\lambda^2 (\varrho)}{a^3} {\Delta}_\bq \bw =
  \frac{\varrho_H \, p'(\varrho)}{a^2 \varrho} {\Delta}_\bq \bP .
\end{equation}
(In the expressions for $\bF$ and $\bPi$, the fields $\varrho$ and
$\bw$ are interchangeable via Eqs.~(\ref{momenteq}c,d), but the
resulting different equations are no longer exactly equivalent when
the approximation $\vec{\cal R}=\b0$ is used.) This term can be
simplified with the ``virial equation of state'' $p = \kappa
\varrho^2$, so that the prefactor of the Laplacian is
density--independent. This expression for $p(\varrho)$ is certainly a
good phenomenological model consistent with the results of numerical
simulations (Dom\'\i nguez 2003, Dom\'\i nguez \& Melott 2004). The
results presented in \S\ref{sec:adhesion} also suggest that the precise value of
the polytropic exponent does not alter the qualitative behavior in
high--density regions. With this simplification, Eq.~(\ref{P_approx})
becomes:
\begin{equation}
  \label{P_approxEJN}
  \ddot{\bP} + 2 H \dot{\bP} - 4 \pi G \varrho_H \bP 
  = \frac{2 \kappa \varrho_H}{a^2} {\Delta}_\bq \bP .
\end{equation}
This equation, being non--linear in Eulerian and in Lagrangian space,
belongs to the class of hybrid Lagrangian/Eulerian models mentioned in
\S\ref{sec:exactgeneral}. As such, the approximation may be viewed --- in
contrast to the Lagrangian perturbation scheme --- in closer
correspondence to the former `adhesion approximation'.  By
construction, this approximation reduces to the full Lagrangian linear
part of the `dust model', Eq.~(\ref{pepsidust}), when the correction
to `dust' is dropped.
Upon linearizing the Eulerian Laplacian, we recover the
Lagrangian linear approximation, Eq.~(\ref{pepsi}). 
(In other words, our non--perturbative approximation
extrapolates the Lagrangian linear model by 
the replacement $\Delta_{\bX} \rightarrow 
\Delta_{\bX + \bP}$.)
Linearization of the convective non--linearities yields the Eulerian
linear approximation, Eq.~(\ref{deltalinear}) (for this, note that linearization
of Eq.~(\ref{rho_general}) gives $\delta = - \nabla_\b0 \cdot \bP$).

\vspace{5pt}

In conclusion, we propose the approximation~(\ref{w_approx}) as a
first extrapolation into the non--perturbative regime of the Eulerian
and Lagrangian linear approximations without resort to the
constraining assumption of parallelism. It should be viewed as a
simplification of the exact Eqs.~(\ref{momenteq}) in the sense that
the approximation yields a local equation, which should facilitate the
theoretical analysis. An attempt to solve the approximate equation
could be successful, but lies beyond the scope of the present work.

\subsection{Regularization of singularities}
\label{sec:regular}

A point one would like to be able to prove rigorously beyond the
argument offered in \S\ref{sec:adhesion} is whether the `EJN and SSE
models' do indeed prevent the formation of singularities given {\em
  initial conditions of cosmological relevance}. 
The absence of singularities when initially smooth data are propagated by the
Vlasov--Poisson system has been proved mathematically 
(Lions \& Perthame 1991; Schaeffer 1991; Pfaffelmoser 1992; Rein \& Rendall 1994).
The original
`adhesion model' reduces to the 3D Burgers' equation
(Eq.~(\ref{burgers2}) with constant $\mu$), which has the unusual
property of being solvable analytically. From there, one can demonstrate
that no singularity arises for any sufficiently smooth initial
conditions (see, e.g., Frisch and Bec 2001).
For other models this is a very difficult question (e.g.,
it is still an open problem for the 3D incompressible
Navier--Stokes equations (Frisch 1995, Sec.~9.3)), and here we can only
provide some general remarks.

Starting from Eqs.~(\ref{momenteq}, \ref{nablau}), one can derive the
following evolution equations for the density $\varrho$ and the
peculiar--expansion rate $\theta$:
\begin{subequations}
  \label{raycha}
  \begin{equation}
    \dot{\varrho} = - 3 H \varrho - \frac{1}{a} \varrho \, \theta ,
  \end{equation}
  \begin{align}
    \dot{\theta} = - 2 H \theta - \frac{1}{3} \theta^2 & - \sigma_{ij} \sigma_{ji} +
    \frac{1}{2} \omega^2 - 4\pi G a(\varrho-\varrho_H) \nonumber \\
    & \mbox{} + \frac{1}{a} \nabla_\bq \cdot \left(\frac{1}{\varrho} \bF - \frac{1}{a \varrho} 
    \nabla_\bq \cdot \bPi \right) .
  \end{align}
\end{subequations}
The continuity equation~(\ref{raycha}a) can be integrated in
Lagrangian coordinates:
\begin{equation}
 \label{lagrrho}
  \varrho(\bX, t) = \varrho(\bX, t_\init) \, \left[ \frac{a_\init}{a} \right]^3 
  \exp \left[- \int_{t_\init}^t \!\!\! d\tau \; \frac{\theta (\bX, \tau)}{a(\tau)} \right].
\end{equation}
Starting from bounded initial conditions, a singularity in the density
field can occur only when 
$\theta \rightarrow -\infty$ at some finite time\footnote{In this work we are
  not concerned with more general (non--Lagrangian) singularities in the
  velocity--gradient field, such as those that may occur in
  incompressible flows.}. From Eq.~(\ref{raycha}b) we
see that only three terms can actually prevent $\theta$ from becoming
too negative: the background expansion, the vorticity, and perhaps the
model--dependent correction to `dust'. In particular, for the `dust
model' with vanishing initial vorticity, only the term linear in $\theta$
in Eq.~(\ref{raycha}b) can oppose collapse, 
so that a density singularity is unavoidable provided there is an initial overdensity.

For the `EJN model', the correction to `dust' reads $-\nabla_\bq \cdot
(\varrho^{-1} \nabla_\bq p)$, where $p$ is modelled as a
density--dependent pressure, so that Eqs.~(\ref{EJN}) have the
structure of the equations of inviscid fluid dynamics. In the absence
of self--gravity, it is known that the solution becomes
non--differentiable for a generic class of smooth initial conditions,
so that the Lagrangian--to--Eulerian map $\bq(\bX,t)$ ceases to be
defined. The map can remain uni--valued if we allow for shocks
(moving discontinuities in the fields): this is called a {\em weak}
solution, and they have to be understood as solutions of the
differential equations~(\ref{EJN}) with an additional viscous term
$\mu \Delta_\bq \bu$ in the limit $\mu \to 0^+$, in the same way as we
discussed for the adhesion model\footnote{The weak solutions can be
  interpreted equivalently as solutions of the {\em integral} form of
  the balance equations for mass and momentum.}.  The presence of
self--gravity is not expected to alter this conclusion; the question
is rather whether the initial conditions of cosmological interest
belong to the class of initial conditions inducing shock formation.
The reason for the emergence of shocks can be ultimately traced back
to the fact that density and peculiar--velocity are independent variables: when
convection by the peculiar--velocity field occurs at such a high rate that the
density gradients are not enough to build up a counteracting pressure,
a singularity arises. However, in the cosmological context the
peculiar--velocity field becomes ``slaved'' to the density field in the linear
regime (see \S\ref{sec:eulerlin}).
The assumption of parallelism~(\ref{parallelism}) extrapolates this
``slaving'' relationship into the non--linear regime: this allows one to
express the density--dependent correction to `dust' in terms of the
velocity gradient, and it seems sufficient to avoid singularities (see
\S\ref{sec:adhesion}). But we have seen that a departure from the
condition of parallelism is generated during the evolution: an open
question is therefore whether the assumption of parallelism is
qualitatively good or instead the dynamically generated departure from
it can lead to the formation of singularities.

The mathematical properties of the equations underlying the `SSE
model' have been barely studied in the literature. The correction to
`dust' is a complicated function of the fields. Unlike the `EJN
model', the corrections depend explicitly on the gradients of the
peculiar--velocity field too, and the analogy with shock formation in inviscid
fluids is not valid.
Another important difference to the `dust' and the `EJN' models is
that vorticity is generated by the `SSE' corrections to `dust' (see
\S\ref{sec:eulernonlin}), which, as pointed out, tends to oppose
collapse. 
Vreman et al.~(1996, 1997) have
studied numerically the hydrodynamical equations of an ideal gas with
an additional stress tensor given by Eq.~(\ref{SSEcorrections}b) when
the initial condition is a perturbation to a smooth profile.
An instability was discovered for which the `SSE'--term is made
responsible. Vreman et al.~(1996) also undertook a theoretical
analysis of the following simplified model (one--dimensional forced
Burgers equation with a simplified, `SSE'--like correction; $\eta>0$):
\begin{equation}
  \frac{\partial u}{\partial t} + u \frac{\partial u}{\partial x} = 
  \mu \frac{\partial^2 u}{\partial x^2} 
  - \eta \frac{\partial}{\partial x} \left( \frac{\partial u}{\partial x} \right)^2 
  +f(x) .
  \end{equation}
It is found that a sinusoidal profile (which fixes the form of the
forcing $f(x)$ by demanding stationarity) is linearly unstable to
perturbations if $2 \eta > \mu-1$.
Indeed, as the discussion of Eq.~(\ref{SSEwork}) below exemplifies,
it is not true that the `SSE'--corrections will behave like viscous terms
under all conditions. As with the `EJN model', the question is
whether the `SSE model' does indeed prevent the formation of
singularities when the initial conditions are restricted to be of
cosmological relevance.

An issue closely related to this problem is the behavior of the
correction to `dust' concerning the local ``energy budget''.  The
density of mean peculiar--kinetic energy is defined as $K:=(1/2)
\varrho |\bu|^2$. From Eqs.~(\ref{momenteq}) one finds: 
\begin{align}
  \label{eq:K}
  \frac{\partial K}{\partial t} + \frac{1}{a} \nabla_{\bq} \cdot (\bu K) & + 5 H K 
  = \bu \cdot (\varrho \bw + \bF) \\
  & \mbox{} + \frac{1}{a} \bPi : (\nabla_{\bq} \bu) - \frac{1}{a} 
  \nabla_{\bq} \cdot (\bPi \cdot \bu)\;, \nonumber
\end{align}
where $:$ indicates contraction of two indices. The first term on the
right--hand--side represents the work done by the total (mean field and
corrections thereof) gravitational force, the second term is the work
done by the velocity dispersion, and the third term is the
contribution due to the flux of velocity dispersion. In the `EJN model',
Eqs.~(\ref{EJNcorrections}), only velocity dispersion does
work: $\bPi : (\nabla_{\bq} \bu)=p(\varrho) \theta$.
Given that $p>0$, this term
tries to reduce the kinetic energy in (relative to the Hubble expansion)
locally collapsing regions
($\theta < 0$) and to increase it in locally expanding ones 
($\theta > 0$).

In the `SSE model', the sign of the work done by the deviations from
mean field gravity, $\bu \cdot \bF = B L^2 (\bu \nabla_{\bq} \varrho)
: (\nabla_{\bq} \bw)$, depends on the orientation of the tidal tensor
and, like the work done by mean field gravity, is
in general unrelated to whether the fluid is locally compressing or
expanding. On the other hand, the work done by velocity dispersion is
most easily expressed in the basis that diagonalizes the symmetric
part of $(1/a) \partial_i u_j$ (this is $\sigma_{ij} + (1/3) \theta
\, \delta_{ij}$, according to the decomposition~(\ref{nablau})):
\begin{multline}
  \label{SSEwork}
  \frac{1}{a} \bPi : (\nabla_{\bq} \bu) = \frac{1}{a} B L^2 \varrho 
  (\partial_k u_i) (\partial_k u_j) (\partial_i u_j) \\
  = 2 B (a L)^2 \varrho \sum_i s_{(i)}
  \left[ s_{(i)}^2 
    + \frac{1}{4} (\omega^2-\omega_i^2) \right] \;,
\end{multline}
where $s_{(i)}$ are the eigenvalues of the tensor $\sigma_{ij} + (1/3)
\theta \, \delta_{ij}$, satisfying $\sum_i s_{(i)}=\theta$.
The work is the sum of contributions along the three eigendirections:
the contribution of the direction of local compression ($s_{(i)} < 0$)
tends to reduce the kinetic energy, that of the direction of local
expansion ($s_{(i)} > 0$) tends to increase it. The net sign is the
outcome of this competition. In this sense, the `SSE' expression for
the velocity dispersion can be viewed as a ``directional''
generalization of the case of an isotropic pressure like in the `EJN
model'. It must be noticed, however, that this competing effect
between local expansion and compression is effective only in the fully
three--dimensional case: it can be easily checked that in the
restricted cases of one or two--dimensional motion,
expression~(\ref{SSEwork}) has the same sign as $\theta$.

\section{Summary and prospects}
\label{sec:summary}

Starting from the equations of motion for point particles, we have
implemented a smoothing procedure in phase space leading to 
equations for the smoothed density and peculiar--velocity fields in Eulerian space.
These equations are the first ones in an infinite
hierarchy of equations, and approximations to close the hierarchy are
required. A closure approximation corresponds physically to an ansatz
on how the evolution of the filtered fields is coupled to the degrees
of freedom below the filtering scales. In this unified context,
we have addressed three different closures, leading to three different
models proposed earlier in the literature: the `dust model', the `EJN
model', and the `SSE model'. 
The `dust model' describes the evolution of fluid volume elements
under the sole influence of the self--consistently generated
gravitational field. The corrections to `dust' take into account the
dynamical influence of the small--scale degrees of freedom. The `EJN
model' approximates this influence as a density--dependent pressure.
The `SSE model' expresses this influence as a functional of density
and velocity in a systematic expansion.

We have provided a clear explanation of how the corrections to the `dust
model' by both the `EJN closure' and the `SSE closure' lead
to models that are qualitatively equivalent to the `adhesion model'. 
The key hypothesis is the parallelism relation~(\ref{parallelism})
between peculiar--velocity and peculiar--acceleration. This already
holds in Zel'dovich's approximation to the exact `dust' evolution ---
the assumption underlying the `adhesion approximation' consists of
extrapolating it to also evaluate the
corrections to `dust'.

We have shown that within both the Eulerian and Lagrangian perturbation
frameworks, the corrections to `dust' violate this assumption in general.
Going beyond the phenomenology of the `adhesion model' therefore requires
relaxing the parallelism assumption. When this is
done, differences between the `EJN model' and the `SSE model' appear:
deviations from the parallelism relationship arise at linear (both
Eulerian and Lagrangian) perturbative order in the `EJN model', but
not in the `SSE model'; vorticity is generated by the `SSE'--corrections
to `dust', but not by the `EJN'--corrections.

The non--perturbative theoretical investigation raises considerable
difficulties. We suggested the plane--symmetric case as the ``exact body'' of
the non--perturbative equations that may serve as a starting point for 
a deeper understanding of the presented models. 
This case can be reduced to the study of a single partial
differential equation for the displacement field, offering the possibility to
study the `EJN' and `SSE' models beyond any approximation.
This simplification misses some features that are expected to be of
relevance for the detailed inner structure of collapsing high--density
regions (e.g., generation of vorticity or the fact that velocity and
gravity have different directions). 
We suggested a new non--perturbative approximation, which 
yields a hybrid Eulerian/Lagrangian partial differential equation for the displacement field
that contains 
as limits (i) the Eulerian and Lagrangian linear approximations, 
and (ii) the standard `adhesion approximation' if one 
imposes the assumption of parallelism of peculiar--velocity and \mbox{--acceleration}.
When particularized to the `EJN model', the `virial equation of state' 
$p \propto \varrho^2$ provides substantial simplifications and, at the same time, is 
a reasonable choice. 
We argued
how this approximation can be viewed as the next step beyond the
parallelism assumption required by the `adhesion model'. 
Finally, we provided a qualitative discussion of the exact `EJN'
and `SSE' models concerning the formation of singularities.

An issue we have not addressed in this work is the choice of the
smoothing scale $L$. This must be guided by the expected applicability
of the corresponding closure approximations, in turn likely related to
the initial conditions. The investigation of this point seems to
require first the ability to extract predictions from the new models
in order to validate them with respect to observations and 
viable N--body simulations.

Further exploration and numerical realization of the models
presented will hopefully lead us not only to approaches capable
of following large--scale structure into the non--linear regime, but
will also offer an interesting way of understanding the role of
previrialization (Peebles \& Groth 1976, Davis \& Peebles 1977,
Peebles 1990, \L okas et al. 1996), and the onset of virialization
in connection with galaxy and galaxy cluster formation. Furthermore
such models can be employed to improve generic inhomogeneous
collapse models (Buchert et al. 2000; Kerscher et al. 2001), and to
enhance the power of reconstructions of initial data from present--day
density and peculiar--velocity fields 
(e.g., Croft \& Gazta{\~n}aga 1997; Brenier et al. 2003; Mohayaee et al. 2003 
for recent efforts). In the latter context, the proposed non--perturbative approximation,
based on the relation $\bw \propto \bP$, Eq.~(\ref{w_approx}), may provide an
alternative to the assumption of parallelism of peculiar--acceleration and
peculiar--velocity, $\bw \propto \bu$, Eq.~(\ref{parallelism}).

\begin{acknowledgements}
  
  This work was motivated by an exciting workshop at Observatoire de
  la C\^ote d'Azur (OCA) on the present subject with financial support
  by OCA, the French Ministry of Education, the Programme National de
  Cosmologie and the Laboratoire G.D. Cassini. Particular thanks
  to Uriel Frisch for organizing this workshop. Fruitful
  discussions were held with him, Erik Aurell, St\'ephane Colombi,
  Jos\'e Gaite, Sergei Gurbatov, Roman Juszkiewicz, Roya Mohayaee,
  Sergei Shandarin, Andrei Sobolevski, Alexei Starobinsky, Takayuki Tatekawa, Roland
  Triay, Patrick Valageas and Barbara Villone.  TB acknowledges
  support by the ``Sonderforschungsbereich SFB 375 f\"ur
  Astro--Teilchenphysik der Deutschen Forschungsgemeinschaft'', and by
  CERN, Geneva, where a preliminary manuscript was written during a
  visit of TB in 1998. AD acknowledges financial support by the
  project ``Formaci\'on de estructuras en astrof\'\i sica y
  cosmolog\'\i a'' (BFM2002--01014) of the Spanish government.
\end{acknowledgements}

\section*{References}

\def\refitem{\par\noindent\hangindent\parindent\hangafter1}

\refitem
Adler, S., \& Buchert, T.\ 1999, A\&A, 343, 317

\refitem 
Antonov, N.V.\ 2004, Phys.\ Rev.\ Lett., 92, 161101

\refitem
Aurell, E., Fanelli, D., \& Gurbatov, S.N., Moshkov, A.Yu.\ 2003, 
Physica D, 186, 171 

\refitem
Bagla, J.S., \& Padmanabhan, T.\ 1994, MNRAS, 266, 227

\refitem
Barbero, F., Dom{\'\i}nguez, A., Goldman, T., \& 
P\'erez--Mercader, J.\ 1997, Europhys.\ Lett., 38, 637

\refitem
Barrow, J.D., \& Maartens, R.\ 1999, Phys.\ Rev.\ D, 59, 043502

\refitem
Berera, A., \& Fang, L.-Z.\ 1994, Phys.\ Rev.\ Lett., 72, 458

\refitem
Bernardeau, F., Colombi, S., Gazta\~naga, E., \& Scoccimarro, R.\ 2002, 
Phys. Rep., 367, 1

\refitem
Bharadwaj, S.\ 1996, ApJ, 472, 1

\refitem
Bildhauer, S., \& Buchert, T.\ 1991, Prog.\ Theor.\ Phys., 86, 653

\refitem
Bildhauer, S., Buchert, T., \& Kasai, M.\ 1992, A\&A, 263, 23 

\refitem
Binney, J., \& Tremaine, S.\ 1987 Galactic Dynamics 
(Princeton Univ.\ Press)

\refitem
Bouchet, F.R., Juszkiewicz, R., Colombi, S., \& Pellat, R.\ 1992,
ApJ, 394, L5

\refitem
Bouchet, F.R., Colombi, S., Hivon, E., \& Juszkiewicz, R.\ 1995,
A\&A, 296, 575

\refitem
Brainerd, T.G., Scherrer, R.J., \& Villumsen, J.V.\ 1993,
ApJ, 418, 570

\refitem
Brenier, Y., Frisch, U., H\'enon, M., Loeper, G., Matarrese, 
S., Mohayaee, R., \& Sobolevskii, A.\ 2003, MNRAS, 346, 501

\refitem
Buchert, T.\ 1989, A\&A, 223, 9

\refitem
Buchert, T.\ 1992, MNRAS, 254, 729

\refitem
Buchert, T.\ 1993a Habilitation Thesis
(Ludwig--Maximilians--Universit\"at M\"unchen; link on ADS)  

\refitem
Buchert, T.\ 1993b, A\&A, 267, L51 

\refitem
Buchert, T.\ 1994, MNRAS, 267, 811 

\refitem
Buchert, T.\ 1996 in Proc.\ IOP `Enrico Fermi', Course CXXXII 
(Dark Matter in the Universe), Varenna 1995, 
ed.\ S.\ Bonometto, J.\ Primack, \& A.\ Provenzale 
(IOS Press Amsterdam), 543

\refitem
Buchert, T.\ 1999 in Proc.\ MPG--CAS Workshop, Shanghai 1998 
({\tt astro--ph/9901002})
 
\refitem
Buchert, T., \& Bartelmann, M.\ 1991, A\&A, 251, 389

\refitem
Buchert, T., \& Dom\'\i nguez, A.\ 1998, A\&A, 335, 395

\refitem
Buchert, T., Dom\'\i nguez, A., \& P\'erez--Mercader, J.\ 1999,
A\&A, 349, 343

\refitem
Buchert, T., \& Ehlers, J.\ 1993, MNRAS, 264, 375

\refitem
Buchert, T., \& Ehlers, J.\ 1997, A\&A, 320, 1

\refitem
Buchert, T., Karakatsanis, G., Klaffl, R., \& Schiller, P.\ 1997, 
A\&A, 318, 1

\refitem
Buchert, T., Kerscher, M., \& Sicka, C.\ 2000,
Phys.\ Rev.\ D, 62, 043525-1-21

\refitem
Buchert, T., Melott, A.L., \& Wei{\ss}, A.G.\ 1994, 
A\&A, 288, 349

\refitem
Chandrasekhar, S.\ 1967 An introduction to the study of 
stellar structure (Dover Pub.)

\refitem
Chandrasekhar, S., \& Lee, E.S.\ 1968, MNRAS, 139, 135

\refitem
Chapman, S., \& Cowling, T.G.\ 1991 The Mathematical Theory of 
  Non-uniform Gases (Cambridge Univ.\ Press)

\refitem
Chernin, A.D., Nagirner, D.I., \& Starikova, S.V.\ 2003, A\&A,
399, 19

\refitem
Clark, R.A., Ferziger, J.H., \& Reynolds, W.C.\ 1979, 
J.\ Fluid Mech., 91, 1 

\refitem
Coles, P., Melott, A.L., \& Shandarin, S.F.\ 1993, MNRAS, 260, 765

\refitem
Croft, R.A.C., \& Gazta{\~n}aga, E.\ 1997, MNRAS, 285, 793

\refitem
Davis, M., \& Peebles, P.J.E.\ 1977, ApJS, 34, 425

\refitem
Dom\'\i nguez, A.\ 2000, Phys.\ Rev.\ D, 62, 103501

\refitem
Dom\'\i nguez, A.\ 2002, MNRAS, 334, 435

\refitem
Dom\'\i nguez, A.\ 2003, Astron.\ Nachr., 324, 560

\refitem
Dom\'\i nguez, A., \& Gaite, J.\ 2001, Europhys.\ Lett., 55, 458

\refitem
Dom\'\i nguez, A., Hochberg, D., Mart{\'\i}n--Garc{\'\i}a, J.M., 
P\'erez--Mercader, J., \& Schulman, L.\ 1999, A\&A, 344, 27; Erratum: 
A\&A, 363, 373

\refitem
Dom\'\i nguez, A., \& Melott, A.L.\ 2004, A\&A, 419, 425

\refitem
Doroshkevich, A.G., Kotok, E.V., Novikov, I.D., Poludov, A.N., 
Shandarin, S.F., \& Sigov, Yu.S.\ 1980, MNRAS, 192, 321 

\refitem
Ehlers, J., \& Buchert, T.\ 1997, Gen.\ Rel.\ Grav., 29, 733

\refitem
Ehlers, J., \& Rienstra, W.\ 1969, ApJ, 155, 105 

\refitem
Fanelli, D., \& Aurell, E.\ 2002, A\&A, 395, 399

\refitem
Frisch, U.\ 1995 Turbulence (Cambridge Univ. Press)

\refitem
Frisch, U., \& Bec, J.\ 2001 in New Trends in Turbulence 
(Les Houches 2000), 
ed.\ M.\ Lesieur, A.\ Yaglom \& F.\ David 
(Springer EDP-Sciences), 341 ({\tt nlin.CD/0012033})

\refitem
G\"otz, G.\ 1988, Class.\ Quant.\ Grav., 5, 743

\refitem
Gorini, V., Kamenshchik, A., Moschella, U., 
\& Pasquier, V.\ 2004, {\tt gr-qc/0403062}

\refitem
Gouda, N., \& Nakamura, T.\ 1988, Prog.\ Theor.\ Phys., 79, 765

\refitem
Gouda, N., \& Nakamura, T.\ 1989, Prog.\ Theor.\ Phys., 81, 633

\refitem 
Gurbatov, S.N., Malakhov, A.N., \& Saichev, A.I.\ 1991 
Nonlinear Random Waves 
and Turbulence in Nondispersive Media (Manchester Univ. Press)

\refitem
Gurbatov, S.N., Saichev, A.I., \& Yakushkin, I.G.\ 1983, 
Sov.\ Phys.\ Usp., 26, 857

\refitem
Gurbatov, S.N., Saichev, A.I., \& Shandarin, S.F.\ 1985, 
Sov.\ Phys.\ Dokl., 30, 921

\refitem
Gurbatov, S.N., Saichev, A.I., \& Shandarin, S.F.\ 1989, 
MNRAS, 236, 385

\refitem
Gurevich, A.V., \& Zybin, K.P.\ 1995, Sov.\ Phys.\ Usp., 38(7), 687

\refitem
Hamana, T.\ 1998, ApJ, 507, L1

\refitem
Haubold, H.J., Mathai, A.M., \& M\"ucket, J.P.\ 1991, 
Astron.\ Nachr., 312, 1

\refitem
Jones, B.J.T.\ 1996, in Proc.\ IOP `Enrico Fermi', Course CXXXII 
(Dark Matter in the Universe), Varenna 1995, 
ed.\ S.\ Bonometto, J.\ Primack, \& A.\ Provenzale 
(IOS Press Amsterdam), 661

\refitem
Kerscher, M., Buchert, T., \& Futamase, T.\ 2001, ApJ, 558, L79

\refitem
Klimas, A.J.\ 1987, J.\ Comput.\ Phys., 68, 202

\refitem
Kofman, L.A.\ 1991 in IUAP Proceedings on Nucleosynthesis in the 
Universe, ed.\ K.\ Sato (Dordrecht: Kluwer), 495  

\refitem
Kofman, L.A., Bertschinger, E., Dekel, A., Gelb, J.M., 
\& Nusser, A.\ 1994, ApJ, 420, 44

\refitem
Kofman, L.A., Pogosyan, D.Yu., \& Shandarin, S.F.\ 1990, 
MNRAS, 242, 200

\refitem
Kofman, L.A., Pogosyan, D.Yu., Shandarin, S.F., \& Melott, A.L.\ 1992, 
ApJ, 393, 437

\refitem
Lions, P.L., \& Perthame, B.\ 1991, Invent.\ Math., 105, 415

\refitem
\L okas, E.L., Juskiewicz, R., Bouchet, F.R., 
\& Hivon, E.\ 1996, ApJ, 467, 1

\refitem
Lynden-Bell, D.\ 1967, MNRAS, 136, 101

\refitem
Ma, C.--P., \& Bertschinger, E.\ 2004, ApJ, 612, 28

\refitem
Maartens, R., Triginer, J., \& Matravers, D.R.\ 1999, 
Phys.\ Rev.\ D, 60, 103503

\refitem
Matarrese, S., Lucchin, F., Moscardini, L., \& S\'aez, D.\ 1992,
MNRAS, 259, 437

\refitem
Matarrese, S., \& Mohayaee, R.\ 2002, MNRAS, 329, 37

\refitem
Melott, A.L.\ 1994, ApJ, 426, L19 

\refitem
Melott, A.L, Buchert, T., \& Wei{\ss}, A.G.\ 1995, A\&A, 294, 345

\refitem
Melott, A.L., Shandarin, S.F., Splinter, R.J., \& Suto, Y.\ 1997, 
ApJ, 479, L79

\refitem
Menci, N.\ 2002, MNRAS, 330, 907

\refitem
Morita, M., \& Tatekawa, T.\ 2001, MNRAS, 328, 815

\refitem
Mohayaee, R., Frisch, U., Matarrese, S., \& Sobolevskii, A 2003, 
A\&A, 406, 393

\refitem
Padmanabhan, T.\ 1996, MNRAS, 278, L29

\refitem
Pauls, J.L., \& Melott, A.L.\ 1995, MNRAS, 274, 99

\refitem
Peebles, P.J.E.\ 1980 The Large--scale Structure of the Universe
(Princeton Univ.\ Press)

\refitem
Peebles, P.J.E.\ 1990, ApJ, 365, 27

\refitem
Peebles, P.J.E., \& Groth, E.J.\ 1976, A\&A, 53, 131

\refitem
Pfaffelmoser, K.\ 1992, J.\ Diff.\ Eqs., 95, 281

\refitem
Pope, S.B.\ 2000 Turbulent Flows (Cambridge Univ.\ Press)

\refitem
Rein, G., \& Rendall, A.D.\ 1994 Arch.\ Rational Mech.\ Analysis 126, 183

\refitem 
Ribeiro, A.L.B., \& Peixoto de Faria, J.G.\ 2005, 
Phys.\ Rev.\ D, 71, 067302

\refitem
Sahni, V., \& Coles, P.\ 1995, Phys.\ Rep., 262, 1

\refitem
Sahni, V., \& Sathyaprakash, B.S., \& Shandarin, S.F.\ 1994,
ApJ, 431, 20

\refitem
Sathyaprakash, B.S., Sahni, V., Munshi, D., Pogosyan, D., 
\& Melott, A.L.\ 1995, MNRAS, 275, 463

\refitem
Schaeffer, J.\ 1991, Part.\ Diff.\ Eqns., 16, 1313

\refitem
Serrin, I.\ 1959 in Encyclopedia of Physics Vol.\ III.1, 
ed.\ S.\ Fl\"ugge (Berlin: Springer)

\refitem
Shandarin, S.F., \& Zel'dovich, Ya.B.\ 1989, Rev.\ Mod.\ Phys., 61, 185

\refitem
Shu, F.H.\ 1978, ApJ, 225, 83

\refitem
Splinter, R.J., Melott, A.L., Shandarin, S.F., \& Suto, Y.\ 1998, 
ApJ, 497, 38

\refitem
Spohn, H.\ 1991 Large Scale Dynamics of Interacting Particles
(Springer--Verlag)

\refitem
Tatekawa, T., Suda, M., Maeda, K., Morita, M., \& Anzai, H.\ 2002, 
Phys.\ Rev.\ D, 66, 064014

\refitem
Tatekawa, T.\ 2004a, Phys.\ Rev.\ D, 69, 084020

\refitem
Tatekawa, T.\ 2004b, Phys.\ Rev.\ D, 70, 064010
 
\refitem
Tatekawa, T.\ 2005, Phys.\ Rev.\ D, 71, 044024

\refitem
Tr\"umper, M.\ 1967, Z.\ Astrophys., 66, 215
 
\refitem
Valageas, P.\ 2001, A\&A, 379, 8

\refitem
Valageas, P.\ 2002, A\&A, 382, 477

\refitem
Vergassola, M., Dubrulle, B., Frisch, U., \& Noullez, A.\ 1994, 
A\&A, 289, 325

\refitem
Vreman, B., Geurts, B., \& Kuerten, H.\ 1996, 
Theor.\ Comp.\ Fluid Dyn., 8, 309

\refitem
Vreman, B., Geurts, B., \& Kuerten, H.\ 1997, 
J.\ Fluid Mech., 339, 357

\refitem
Weinberg, D.H., \& Gunn, J.\ 1990, MNRAS, 247, 260

\refitem
Wei{\ss}, A.G., Gottl\"ober, S., \& Buchert, T.\ 1998, MNRAS, 278, 953

\refitem
Yano, T., \& Gouda, N.\ 1998, ApJS, 118, 267

\refitem
Yano, T., Koyama, H., Buchert, T., \& Gouda, N.\ 2003, ApJS, 151, 185

\refitem
Zel'dovich, Ya.B.\ 1970, A\&A, 5, 84

\refitem
Zel'dovich, Ya.B.\ 1973, Astrophysics, 6, 164

\refitem
Zel'dovich, Ya.B., \& Shandarin, S.F.\ 1982, 
Sov.\ Astron.\ Lett., 8(3), 139

\appendix

\section{On the transformation to comoving coordinates}
\label{app:comoving}

To derive Eqs.~(\ref{momenteq}) we first introduced comoving
coordinates, Eqs.~(\ref{comovingtrafo}), and then the coarse--graining procedure was applied.
It is legitimate to ask whether the order can be exchanged, i.e., what
are the transformations that lead from Eqs.~(\ref{hydroeq}) to
Eqs.~(\ref{momenteq})? They can be obtained easily by inserting the
transformations~(\ref{comovingtrafo}) into the
definitions~(\ref{defs}, \ref{corrections}) and using the
definitions~(\ref{hydrodefs}, \ref{hydrocorrections}).
One gets (here we write explicitly the dependence on the smoothing
length):
\begin{subequations}
  \label{comovingtrafo2}
  \begin{equation}
    \bq = \frac{1}{a} \bx , \qquad L = \frac{1}{a} {\cal L} \;,
  \end{equation}
  \begin{equation}
    \varrho (\bq, t; L) = \rho (\bx, t; {\cal L}) \;, 
  \end{equation}
  \begin{equation}
    \bu (\bq, t; L) = \bv (\bx, t; {\cal L}) - 
    H \vec{R} (\bx, t; {\cal L}) \;,
  \end{equation}
  \begin{equation}
    \bw (\bq, t; L) = \bg (\bx, t; {\cal L}) + 
    \frac{4 \pi G \varrho_H - \Lambda}{3} \bx \;, 
  \end{equation}
  \begin{equation}
    \bF (\bq, t; L) = \vec{\cal F} (\bx, t; {\cal L}) + 
    \frac{4 \pi G \varrho_H - \Lambda}{3} (\vec{R}-\bx) \rho (\bx, t; {\cal L})\;, 
  \end{equation}
  \begin{equation}
    \Pi_{ij} (\bq, t; L) = {\cal P}_{ij} (\bx, t; {\cal L}) - 
    H [{\cal D}_{ij} (\bx, t; {\cal L}) + {\cal D}_{ji} (\bx, t; {\cal L})] 
   \nonumber
   \end{equation}
   \begin{equation}
   + H^2 {\cal I}_{ij} (\bx, t; {\cal L}) \;. 
  \end{equation}
\end{subequations}
The following new quantities had to be introduced:
\begin{subequations}
  \begin{equation}
    \rho \vec{R} (\bx, t; {\cal L}) := \frac{m}{{\cal L}^3} \sum_\alpha 
    W \left(\frac{\bx-\bx^{(\alpha)}}{{\cal L}} \right) \bx^{(\alpha)} ,
  \end{equation}
  \begin{eqnarray}
    {\cal D}_{ij} (\bx, t; {\cal L}) := \frac{m}{{\cal L}^3} \sum_\alpha 
    W \left(\frac{\bx-\bx^{(\alpha)}}{{\cal L}} \right)
    v_i^{(\alpha)} x_j^{(\alpha)} \nonumber\\ 
    - \rho v_i R_j (\bx, t; {\cal L}) ,
  \end{eqnarray}
  \begin{eqnarray}
    {\cal I}_{ij} (\bx, t; {\cal L}) := \frac{m}{{\cal L}^3} \sum_\alpha 
    W \left(\frac{\bx-\bx^{(\alpha)}}{{\cal L}} \right)
    x_i^{(\alpha)} x_j^{(\alpha)} \nonumber\\
   - \rho R_i R_j (\bx, t; {\cal L}) .
  \end{eqnarray}
\end{subequations}
$\vec{R}$ is the center--of--mass position of the subsystem defined by
the smoothing window, and $\vec{\cal I}$ is the inertia tensor of this
same subsystem with respect to $\vec{R}$. Concerning $\vec{\cal D}$, we note that 
the antisymmetrized combination $\epsilon_{ijk} {\cal D}_{kj}$ is the
angular momentum of this subsystem with respect to $\vec{R}$; 
the transformation~(\ref{comovingtrafo2}f)
involves only the {\em symmetrized} tensor.
Therefore, to obtain a self--contained set of
transformations, new dynamical equations for $\vec{R}$, $\vec{\cal
  I}$, $\vec{\cal D}$ would be required, including also further
approximations to close the corresponding hierarchy. The simplest one
is provided by the `dust model': the small--scale degrees of freedom are
neglected altogether as irrelevant for the dynamics, so that
$\vec{R}=\bx$, $\vec{\cal I}=\b0$, $\vec{\cal D}=\b0$, and the
transformation~(\ref{comovingtrafo2}) reduces to the standard
one (e.g., Peebles 1980).

Compared to this `standard transformation',
transformation~(\ref{comovingtrafo2}) differs in two points: (i) the
additional transformation rules~(\ref{comovingtrafo2}e,f) involving
the fields $\bF$ and $\bPi$, and (ii) the transformation between
$\bu$ and $\bv$, because in general $\vec{R} (\bx, t)
\neq \bx$. When the particle distribution looks
homogeneous at all cosmological scales, the deviation $\vec{R}-\bx$ is
irrelevant. In the highly non--linear regime of large inhomogeneities, one
expects $|\vec{R} -\bx| \sim a L$, but then $|\bv|, |\bu| \gg H a L$ nevertheless, and 
the cosmological expansion is also irrelevant (Dom\'\i nguez \& Gaite 2001). The
difference may be important in the most interesting case when the
structure is beginning to enter the non--linear regime, $|\bv|, |\bu|
\sim H a L$.

Although the change to comoving coordinates does not affect the
physical content of the resulting equivalent equations,
(\ref{hydroeq}) or (\ref{momenteq}), this equivalence can be broken after
approximations are made. It seems more straightforward to ``subtract''
the known background cosmological expansion before performing any
smoothing/averaging, as we did to derive Eqs.~(\ref{momenteq}), so
that the approximations are concerned only with the ``unknown'' part
of the evolution of the inhomogeneities. However, the above remarks
show that this procedure ``hides'' additional physics and
requires a careful definition of ``background'' at the
coarse--graining scale, which is only implicit by coarse--graining in
the pre--defined comoving space.

\section{Derivation of the `Small--Size--Expansion (SSE)'}
\label{app:sse}

In this Appendix we recall the derivation of the `SSE'
closure~(\ref{SSEcorrections}) (Dom\'\i nguez 2000). 
In analogy with definitions~(\ref{defs}), one can formally
define
microscopic fields by integrating the Klimontovich density~(\ref{f_K})
over the velocity variable\footnote{As with Eqs.~(\ref{klim}), the
  formal definitions of the fields $\bu_{mic}(\bq)$, $\bw_{mic}(\bq)$
  can be regularized by smearing Dirac's delta function.}:
\begin{subequations}
  \begin{eqnarray}
    \varrho_{mic} (\bq, t) & := & m \int d\bu \; 
    f_K (a \bq, \bu-\dot{a} \bq, t) \nonumber \\
    & = & \frac{m}{a^3} \sum_\alpha 
    \delta (\bq - \bq^{(\alpha)}) \;,
  \end{eqnarray}
  \begin{eqnarray}
    \varrho_{mic} \bu_{mic} (\bq, t) & := & m \int d\bu \; \bu \, 
    f_K (a \bq, \bu-\dot{a} \bq, t) \nonumber \\
    & = & \frac{m}{a^3} \sum_\alpha \bu^{(\alpha)} \, 
    \delta (\bq - \bq^{(\alpha)}) \;,
  \end{eqnarray}
  \begin{eqnarray}
    \label{wmic}
    \varrho_{mic} \bw_{mic} (\bq, t) & := & m \int d\bu \; \bw (\bq, t) \, 
    f_K (a \bq, \bu-\dot{a} \bq, t) \nonumber \\
    & = & \frac{m}{a^3} \sum_\alpha \bw^{(\alpha)} \, 
    \delta (\bq - \bq^{(\alpha)}) \;,
  \end{eqnarray}
\end{subequations}
already expressed in comoving coordinates. 
In Eq.~(\ref{wmic}), $\bw(\bq, t) := \bg(a\bq, t) + (4 \pi G \varrho_H(t) - 
\Lambda) a \bq/3$, with $\bg$ given by Eqs.~(\ref{klim}b). 
Note that $\bw_{mic}(\bq, t) \equiv \bw(\bq,t)$,
so that by virtue of Eqs.~(\ref{klim}b) (when expressed in comoving coordinates), 
we can write:
\begin{equation}
  \label{micropoisson}
  \nabla_\bq \cdot \bw_{mic} = - 4 \pi G a (\varrho_{mic} - \varrho_H) \;, \qquad
  \nabla_\bq \times \bw_{mic} = \b0 \;. 
\end{equation}
In terms of the microscopic fields, definitions~(\ref{defs},
\ref{corrections}) can be rewritten as follows:
\begin{subequations}
  \label{micro}
  \begin{equation}
    \varrho (\bq, t) = \int \frac{d\bq'}{L^3} \; W \left(\frac{\bq -\bq'}{L} 
    \right) \varrho_{mic} (\bq', t)\; ,
  \end{equation}
  \begin{equation}
    \varrho \bar\bu (\bq, t) = \int \frac{d\bq'}{L^3} \; 
    W \left(\frac{\bq -\bq'}{L} \right) 
    \varrho_{mic} \bu_{mic} (\bq', t) \;,
  \end{equation}
  \begin{equation}
    \bF (\bq, t) = \int \frac{d\bq'}{L^3} \; W \left(\frac{\bq -\bq'}{L} \right) 
    \varrho_{mic} \bw_{mic} (\bq', t) - \varrho \bar\bw (\bq, t) \;,
  \end{equation}
  \begin{eqnarray}
    \bPi (\bq, t)  = \int \frac{d\bq'}{L^3} \; W \left(\frac{\bq -\bq'}{L} \right) 
    \varrho_{mic} \bu_{mic} \bu_{mic} (\bq', t)\nonumber\\ 
   -  \varrho \bar\bu \bar\bu (\bq, t) \;,
  \end{eqnarray}
\end{subequations}
where $\bar\bw$ is the mean field strength of the peculiar--gravity,
defined by Eqs.~(\ref{momenteq}c,d).
The introduction of the
{\it mic}--fields is a formal step to express the definitions in terms
of volume integrals over fields.
Notice in particular that $\bu$ is defined by a volume--average of the
peculiar--momentum density (i.e., by a {\em mass--average} of the
velocity) and thus it has the physical meaning of center--of--mass
velocity of the coarsening cell. These fields obey
Eqs.~(\ref{micro}a, b), which can be formally inverted to yield the following
expansion in $L$ (this is most easily derived in Fourier space; this
formal inversion is defined provided the Fourier transform of the
window $W(\cdot)$ has no zeros):
\begin{subequations}
  \label{inversion}
  \begin{equation}
    \varrho_{mic} = \left[ 1 - \frac{1}{2} B \, (L \, \nabla_{\bq})^2 + 
     {\cal O}(L \, \nabla_{\bq})^4 \right] \, \varrho \;,
  \end{equation}
  \begin{equation}
    \bu_{mic} = \left[ 1 - \frac{1}{2} B \, (L \, \nabla_{\bq})^2 - B L^2
      \frac{\nabla_{\bq} \varrho}{\varrho} \cdot \nabla_{\bq} + {\cal O}(L \, 
      \nabla_{\bq})^4 \right] \, \bar\bu \;,
  \end{equation}
\end{subequations}
with 
\begin{equation}
  \label{eq:B}
  \int d \vec{z} \; \vec{z} \, W(\vec{z}) = \b0 \;\;, \qquad
  B := \frac{1}{3} \int d \vec{z} \; z^2 \, W(\vec{z}) \;.
\end{equation}
These expansions show how the microscopic fields can be recovered
by taking into account finer and finer details in the spatial
distribution of the coarse--grained fields. To derive
Eq.~(\ref{SSEcorrections}b), one inserts Eqs.~(\ref{inversion}) into
the definition~(\ref{micro}d) encompassing the coupling to the
small--scale kinetic degrees of freedom:
\begin{eqnarray*}
  \bPi (\bq, t) & = &
  \int \frac{d\bq'}{L^3} \; W \left(\frac{\bq -\bq'}{L} \right) 
  \{ \varrho \bar\bu \bar\bu (\bq', t) - \varrho \bar\bu \bar\bu (\bq, t) \\
  & & \mbox{} - \frac{1}{2} B L^2  
  [\bar\bu \nabla_{\bq'}^2 (\varrho \bar\bu) (\bq', t) 
  + (\nabla_{\bq'}^2 \bar\bu) \varrho \bar\bu (\bq', t)  \\
  & & \mbox{} + 2 (\nabla_{\bq'} \varrho) \cdot (\nabla_{\bq'} \bar\bu) \bar\bu (\bq', t)]\} 
  + {\cal O}(L^4) \\
  & = & B L^2 \varrho (\partial_k \bar\bu) (\partial_k \bar\bu) (\bq, t)
  + {\cal O}(L^4) \;,
\end{eqnarray*}
where the fields in the integrand have been expanded about $\bq'=\bq$
consistently to a given order in $L$. The expansion
(\ref{SSEcorrections}a) for the correction to mean field is derived
in the same manner 
(for this, notice that $\bw_{mic}$ is a linear functional of
$\varrho_{mic}$ given by Eqs.~(\ref{micropoisson})).

\section{Dynamical equations of state and corresponding model features}
\label{app:eqstate}

An ingredient of the `EJN' model is an ``equation of
state''\footnote{This notion as well as the names attached to the
  different polytropic models (with the polytropic index $n$ defined through 
 $\gamma=:(n+1)/n$) 
  is borrowed from thermodynamics.
  We emphasize that here we refer to dynamical stresses and
  do not imply thermal equilibrium.}  
$p(\varrho)$ for the trace of the velocity dispersion tensor as
function of the mass density, Eq.~(\ref{EJNcorrections}).  In this
Appendix we summarize the evidence in favor of this assumption and the
physical meaning of some particular functional dependences.

\subsection*{Numerical simulations}

The density--dependence of the trace of the velocity
dispersion~(\ref{corrections}b) was addressed systematically with
N--body simulations of CDM scenarios by Dom\'\i nguez (2003) and
Dom\'\i nguez \& Melott (2004) and can be summarized in a (sometimes
piecewise) polytropic dependence $p \propto \varrho^\gamma$,
which should be understood as an average, because
the numerical data scatter around this dependence.
The  index $\gamma$ is always found to lie in the range $1
\leq \gamma \leq 2$. The more power there is in the small scales of
the initial mass distribution relative to the large scales, the
smaller is $\gamma$.

\subsection*{Special values of $\gamma$}

There are values of the index $\gamma$ which have a
particular significance (see also Buchert et al.~1999):
\begin{list}{$\bullet$}{\leftmargin=0pt \itemsep=5mm \itemindent=4mm}
\item $\gamma=2$: `Virialized state' 
  
  This model is frequently encountered in our paper. From a
  mathematical point of view, the index $\gamma=2$ entails a
  simplification because the length $\lambda(\varrho)$,
  Eq.~(\ref{lambda}), becomes $\varrho$--independent. Thus, the
  standard `adhesion model' is recovered in this case because the GM
  coefficient $\mu$, Eq.~(\ref{GMcoeff}), is also
  $\varrho$--independent; see also the derivation of
  Eq.~(\ref{P_approxEJN}).
  
  Physically, one can argue on the basis of the virial theorem
  (e.g.~Chandrasekhar \& Lee, 1968): a well--known
  consequence of it is that, if a system dominated by gravity is
  {\em isolated} and in a {\em stationary} state, then the total
  kinetic energy ${\cal U}$ (in the frame of the system's center of
  mass) and the total (gravitational) potential energy ${\cal W}$ are
  related by the equation $2 {\cal U} + {\cal W} = 0$. If our system is
  a coarsening cell of size $\sim L$ defined by the smoothing
  window $W(\cdot)$, we have ${\cal U} \propto p$ and ${\cal W} \sim G
  L^5 \varrho^2$ assuming that ${\cal W}$ can be estimated reliably
  within a mean field approximation (i.e., not too much small--scale
  structure). To the extent that a coarsening cell can be considered
  approximately isolated and stationary, the virial theorem implies $p
  \propto \varrho^2$. The use of this relationship in the evolution
  equation~(\ref{EJN}b) always supposes an approximation, because the
  term $-\nabla p$ represents the flux of momentum by the exchange of
  particles between neighboring coarsening cells, in contradiction with
  the assumption of isolation required to derive the dependence $p
  \propto \varrho^2$.
  
\item $\gamma=1$: `Isothermal state'
  
  In this case, the plane--symmetric problem of the `EJN model', in the absence
  of cosmological expansion, can be mapped exactly to the `Sine--Gordon
  equation' (G\"otz, 1988).
    
  Physically, this dependence implies that the kinetic energy per
  particle, which is proportional to $p/\varrho$, is
  density--independent. In the context of a fluid in thermal
  equilibrium, the reason is that, concerning any single particle, the
  rest of the system behaves like a thermostat.
  
\item $\gamma=5/3$: `Adiabatic state'
 
  Starting from the definition~(\ref{corrections}b), one can compute
  the following evolution equation for the tensor $\Pi_{ij}$: 
    \begin{eqnarray}
    \label{eqPi}
    \frac{\partial \Pi_{ij}}{\partial t} + 5 H \Pi_{ij} & = & 
    - \frac{1}{a} \Pi_{ik} \, (\partial_k u_j) - \frac{1}{a} \Pi_{jk} \, (\partial_k u_i) \\
    & & - \frac{1}{a} \partial_k (u_k \, \Pi_{ij}) -
    \frac{1}{a} \partial_k {\cal L}_{ijk} + {\cal P}_{ij} . \nonumber
  \end{eqnarray}
  In this equation, ${\cal L}_{ijk}$ is the third reduced velocity moment 
  (representing the heat flux in dilute gases) and
  ${\cal P}_{ij}$ involves the departures from mean field. (The
  precise expressions for these two tensors can be found in (Dom\'\i
  nguez 2000)). One may take the closure assumption that these two
  tensors are negligible, meaning physically that corrections to mean
  field are unimportant, and that velocity dispersion is relatively
  small (and with it the ``heat flux'' $\Rightarrow$ ``adiabatic
  evolution''). If one further makes the isotropy ansatz $\Pi_{ij} = p
  \delta_{ij}$, this equation in combination with the continuity
  equation~(\ref{momenteq}a) then yields $p = \kappa \varrho^{5/3}$ and
  the strong kinematical constraint of vanishing shear, $\partial_i
  u_j + \partial_j u_i \propto \delta_{ij}$ (Buchert \& Dom\'\i nguez 1998).
  Note that this implies that the only consistent exact
  solution is spatially homogeneous (Tr\"umper 1967; Ehlers \& Rienstra 1969; 
  the homogeneous model also satisfies `adiabaticity'.)

The coefficient of proportionality $\kappa$ is a local
    function of the Lagrangian coordinates given by the initial data;
    if we assume that, initially, this function is a spatial constant,
    then we are entitled to talk about a ``state'', since the relation
    between dynamical ``pressure'' and mass density is then global.

\item $\gamma=3$: Plane symmetric `adiabatic state' 
  
  In a plane--symmetric configuration, Eq.~(\ref{eqPi}) can be
  integrated under the assumption that ${\cal L}_{ijk}$ and ${\cal
    P}_{ij}$ are negligible without the need for the additional isotropy
  ansatz for $\Pi_{ij}$. Thus, the equation for the single component
  $\Pi_{xx}$ becomes
  \begin{displaymath}
    \frac{\partial \Pi_{xx}}{\partial t} + \frac{1}{a} u \frac{\partial \Pi_{xx}}{\partial x}
    + 5 H \Pi_{xx} = 
    - \frac{3}{a} \Pi_{xx} \, \frac{\partial u}{\partial x} ,
  \end{displaymath}
  from which one obtains $\Pi_{xx} \propto \varrho^3$.
  
  Without regard to this derivation,  in the
  context of the generalized `adhesion models', the $\varrho$--dependence of the GM coefficient
  derived with the `SSE model', Eq.~(\ref{GM_SSE}), can be obtained
  with the `EJN model' and a dependence $p \propto \varrho^3$,
  Eq.~(\ref{GMcoeff}).
  
\item $\gamma=4/3$: `Cosmogenetic state'
  
  If one looks for a solution of Eqs.~(\ref{EJN}) in ``comoving
  hydrostatic equilibrium'', i.e., $\bu=\b0$ but $\nabla \varrho \neq
  \b0$, with an ``equation of state'' $p = \kappa \varrho^\gamma$,
  then one can easily check that $\gamma=4/3$ is the only possible
  index for which a solution exists when $\kappa$ is
  time--independent (Chandrasekhar 1967).

\item $\gamma=-1$: `Chaplygin state'
  
  In our context, this formal case is of interest because the
  particularization of the `EJN model' to a plane--symmetric
  configuration gives a linear equation for the Lagrangian
  displacement field $\bP$ exactly, see \S\ref{sec:exactplane}.
  
  More generally, the index $\gamma=-1$ introduces
  mathematical simplifications in the integration of Euler's equation.
  Recently, the Chaplygin gas has received attention in
  cosmological research as one possible model of dark energy (see
  e.g.~the recent preprint by Gorini {\em et al.} 2004 and
  refs.~therein).
  
\item $\gamma<1$:  
  
  Application of boundary--layer theory to Eqs.~(\ref{burgers2})
  (Dom\'\i nguez 2000) shows that density singularities are
  regularized when $\gamma>1$. On the other hand, a similar analysis
  (Dom\'\i nguez, unpublished) demonstrates that ``pressure'' is too
  weak to avoid the occurrence of singularities when $\gamma<1$.

\end{list}

\end{document}